\documentclass[fleqn,usenatbib]{mnras}

\usepackage{newtxtext,newtxmath}

\usepackage[T1]{fontenc}

\DeclareRobustCommand{\VAN}[3]{#2}
\let\VANthebibliography\thebibliography
\def\thebibliography{\DeclareRobustCommand{\VAN}[3]{##3}\VANthebibliography}

\usepackage{graphicx}	
\usepackage{amsmath}	
\usepackage{comment}
\usepackage{float}
\usepackage{algorithm}
\usepackage{algpseudocode}
\usepackage{hyperref}
\usepackage[normalem]{ulem}
\usepackage{soul}
\usepackage{multirow}
\usepackage{fontawesome}
\newtheorem{Proposition}{Proposition}

\newcommand{\planck}{\emph{Planck} } 
\newcommand{\lb}{LiteBIRD } 
\newcommand{\sosat}{SO-SAT } 

\title[ Foreground Removal with Clustering Methods ]{Improved Galactic Foreground Removal for B-Modes Detection with Clustering Methods}

\author[Giuseppe Puglisi]{Giuseppe Puglisi$^{1,2,3}$,\thanks{E-mail: gpuglisi@lbl.gov}
Gueorgui Mihaylov$^{4}$,
Georgia V. Panopoulou$^{5}$\thanks{Hubble Fellow},
Davide Poletti$^{6,7}$, 
Josquin Errard$^8$,
\newauthor
Paola A.  Puglisi$ ^{10} $,
Giacomo Vianello$^{11} $
\\
$^1${Computational Cosmology Center, Lawrence Berkeley National Laboratory, Berkeley, CA 94720, USA }\\
$^2${Space Sciences Laboratory at University of California, 7 Gauss Way, Berkeley, CA 94720 }\\
$^3${Department of Physics, University of California, Berkeley, CA, USA 94720 }\\
$^{4}$GSK and King's College London, Department of Mathematics , Strand, London UK\\
$^{5}$California Institute of Technology, MC350-17, 1200 East California Boulevard, Pasadena, CA 91125, USA\\
$^6$Universit\'a degli Studi di Milano-Bicocca, Piazza della Scienza 3, 20126 Milano, Italy \\
$^7$ INFN - Sezione di Milano Bicocca, Piazza della Scienza 3, 20126 Milano, Italy \\
$^{8}$ AstroParticule et Cosmologie, Univ. Paris Diderot, CNRS/IN2P3, CEA/Irfu, Obs de Paris, Sorbonne Paris Cit\'e , France\\
$^{9}$Royal Mail, 185 Farringdon Rd, London EC1A 1AA, UK \\
$^{10}$Cape Analytics, 100 W Evelyn Ave UNIT 220, Mountain View, CA 94041, USA  \\ 
}
  
\date{Accepted XXX. Received YYY; in original form ZZZ}

\pubyear{2021}

\begin{document}
\label{firstpage}
\pagerange{\pageref{firstpage}--\pageref{lastpage}}
\maketitle

\begin{abstract}
Characterizing the    sub-mm Galactic emission has become increasingly critical especially in  identifying and removing  its polarized contribution from the  one emitted by the Cosmic Microwave Background (CMB). In this work, we  present a parametric  foreground removal performed onto  sub-patches identified in the celestial sphere by means of spectral clustering. Our approach takes into account efficiently both the geometrical affinity and the similarity induced by the measurements and the accompanying errors.    The optimal partition is then used  to   parametrically separate the Galactic emission encoding thermal dust and synchrotron  from the CMB one applied on two   nominal observations of forthcoming experiments from the ground and from the space.  Moreover, the clustering is performed on tracers  that are different from the data used for component separation,  e.g.  the spectral index maps of  dust and synchrotron. 
Performing the parametric fit singularly on each of the clustering derived regions results in an overall improvement:  both  controlling the bias and the   uncertainties in the CMB $B-$mode recovered maps.  
We finally apply this technique using the map of the number of clouds along the line of sight, $\mathcal{N}_c$,  as estimated from HI emission data and  perform parametric fitting onto patches derived by clustering on this map.  We show that  adopting the  $\mathcal{N}_c$ map as a tracer for the patches related to the thermal dust emission,  results in  reducing the $B-$mode residuals post-component separation. The code is  made publicly available \href{https://github.com/giuspugl/fgcluster}{\faGithub}.

\end{abstract}




\section{Introduction}

In the past few decades, the anisotropies of the Cosmic Microwave Background (CMB) have been measured  to constrain the parameters of the standard cosmological model, the  $\Lambda$-Cold-Dark-Matter ($\Lambda$CDM), with unprecedented precision \citep{2016A&A...594A..13P,planck2018_cosmopars}. In recent years, the polarization of the CMB has received significant attention. The linear polarization of the CMB arises primarily from Thompson scattering of photons with free electrons in the photon-baryon plasma of the epoch of recombination. The linearly polarized emission anisotropies can be decomposed into a curl-less field, commonly referred to as the $E-$modes, and a divergence-less one, known as $B-$modes \citep{1997PhRvL..78.2054S,Hu:1997hv}. The $E-$modes are linked to the primordial scalar perturbations, whereas $B-$modes at the degree scale  can be produced  only by tensor perturbations of the space-time metric  emitted at the \emph{inflationary} era \citep{1981PhRvD..23..347G,1982PhLB..117..175S}.  $B-$modes are expected to be observed at   degree angular scales,  making them an interesting scientific  target to validate several  theoretical  models describing  the early universe as their amplitude, commonly  parametrized by the tensor-to-scalar ratio $r$, is proportional to the energy scale when inflation occurred. 

At smaller angular scales ($\sim$arcmin), $B-$modes can be sourced by the direct distortion of $E-$modes  due to gravitational lensing  of the intervening large scale structure. Lensing $B-$modes have been detected in the past  years by many ground-based experiments,  \citep[e.g.][]{ThePOLARBEARCollaboration2017,pb2019,choi2020atacama,Bianchini_2020}. On the other hand, primordial B-modes have not been detected yet and  the best constraints  have recently been set to $r<0.044, \, 95 \% $ confidence level, by \citet{Tristram_2021}  combining data from the BICEP2/Keck Array and Planck experiments. 

Together with the instrumental sensitivity,   the major challenge to detect primordial $B-$modes  is the polarized  foreground radiation emitted at the same CMB frequencies ($\nu\sim 100$\,GHz)   from our own Galaxy.  Electrons accelerated by spiralling along the Galactic magnetic field lines emit  synchrotron emission   mostly dominating the foreground polarized emission at low frequencies, $\nu< 70 $\,GHz.   On the other hand, thermal dust emission arising from grains aligned with the Galactic magnetic field  dominates the high frequency end of the sub-mm foreground radiation,  $\nu>100 $\,GHz. Polarization from other processes, \citep[e.g. from molecular lines][]{2017MNRAS.469.2982P},  is  expected to be subdominant with respect to that of Galactic synchrotron and dust emission \citep{pldiffuse2015,pldiffuse2018}.

Since the Galactic components have different  frequency dependences, they can be disentangled by means of component separation or foreground cleaning techniques \citep[][]{2014A&A...571A..12P,2016A&A...594A...9P,2016A&A...594A..10P}.   On  one hand, \emph{blind} algorithms, (e.g.  Internal Linear Combination) are usually aimed at removing  emission that is different than the CMB, exploiting either the assumption of statistical independence for most of the sky components \citep[see][]{Remazeilles_2018,2003MNRAS.346.1089D,2007MNRAS.374.1207M} or the maximum entropy principle \citep{2005MNRAS.357..145S}.  On the other hand,  \emph{non-blind}  approaches e.g. internal  template subtraction  or parametric fitting \citep[][]{dunkley09,2008ApJ...676...10E,2008A&A...491..597L,stompor08,2008StMet...5..307B,2006ApJ...648..784H,1992ApJ...396L...7B} are employed in order to distinguish several Galactic components. These come at the cost of making assumptions to model the frequency scaling of all the emission involved, commonly accounting for a very large number of free parameters. 
Moreover, what accentuates the difficulties of foreground removal is the fact that  the properties of the interstellar medium (ISM) change not only \emph{spatially}, i.e. in different locations in the sky \citep{pldiffuse2015,pldiffuse2018} but also along the same line-of-sight \citep{tassis2015,Clark2018,Panopoulou2020}, eventually leading  to  frequency dependencies that are more complex than typically assumed  \citep[][]{Chluba_2017,Pelgrims2021,Mangilli_2021,Azzoni_2021}.  As a result,  algorithms relying on parametric component separation would tend to  reconstruct  CMB maps with large    residual foreground bias  as   the spatial variability of foregrounds is mismodelled by fitting a single parameter across the whole observed sky.

A partition of the sky, where the parameters are fit independently on  multiple regions, results in  a mitigation of   the  bias   but might increase  the statistical uncertainties due to instrumental noise. This is mainly due to the fact that  the fit is   performed on a smaller number  of pixels, encoding   a lower signal-to-noise ratio  (SNR) than the case with all the observed pixels.  Moreover, performing a pixel-by-pixel  foreground removal is clearly unfeasible for experiments involving large sky footprints and/or high resolution because fitting for spectral indices in each sky pixel is a very costly process. 

Recently, several techniques to encompass the spatial variability of the foregrounds have been proposed by decomposing the map into wavelets or needlets \citep{basak2011,Remazeilles_2018,Wagner-Carena2019,2019A&A...623A..21I},   by partitioning the sky into regions (clusters) according to the similarity of foreground properties and their location in the sky \citep{Khatri_2019,Grumitt_2020}. 

In this work,  we do spectral clustering on a proxy of SED properties of the Galactic  foregrounds by considering the spectral parameters and the number of clouds along the line of sight map.  Once the clusters are defined used them for component-separating multi-frequency observations \citep{stompor08,2011PhRvD..84f3005E,PhysRevD.94.083526,PhysRevD.95.043504}.  

The approach is similar to the one shown in \citet{Grumitt_2020} but we choose a different \emph{clustering method} to optimally divide the sky.  In fact, \citet{Grumitt_2020} used the \emph{mean-shift} algorithm \citep{Krzanowski1988} to  locate overdensities in the context of image segmentation. 
  They developed a parametric bayesian component separation algorithm that makes use of a clustering analysis to forecast the accuracy of component separation for a LiteBIRD-like CMB space satellite mission \citep{Sugai_2020}. They  define clusters in a 5-dimensional parameter space: 3-cartesian coordinates of pixels of the sphere, and the spectral indices  $\beta_s$ and $\beta_d$ used to parametrize respectively the emission of synchrotron and dust. However, as noted by the authors, more work is needed to identify the optimal sky templates for this kind of study. An obvious area of improvement is to optimize the definition of sky regions.   The characteristics of   clusters obtained in \citet{Grumitt_2020} were almost homogeneous across the sky: showing  approximately the same size and similar (convex) shapes.  However, the mean-shift cluster size is highly dependent on the choice of the  bandwidth parameter in the definition of the kernels to identify the clusters. Ideally, the clustering should reflect the underlying spatial distribution of the features of interest. More sophisticated algorithms for image segmentation are available for this. In this work, we focus on improving upon the definition of clusters on the sky.

We propose in this work a different algorithm to partition the sky via \emph{spectral clustering} \citep{VonLuxburg2007}.  In this specific application,  the foreground spectral parameter maps are defined over the whole celestial sphere and can thus be mathematically treated as real-valued functions defined on   the  real manifold $S^2$.  In the context of unsupervised image segmentation,   spectral clustering has proved to be a powerful technique able to capture objects with highly non-trivial geometric shapes \citep{Zhang2018,Zelnik-Manor}. 
 Our approach takes into account the degree of similarity inferred both from  geometric positions and from the measurements performed in different  points of the sky.  Being based on an eigen-decomposition, it is less affected by high-dimensionality issues. Furthermore,   we use  the signal-to-noise content of the spectral parameter maps  to allow the pixel similarity  to be informed  by  the intrinsic variability of a given foreground parameter map.

The definition of clusters on a manifold can be divided into two procedures: (i) building  the pixel  similarity accounting for the ``distances'' in a given metric, (ii) finding the eigenspectrum decomposition   for a given set of features. In  Sect.~\ref{sec:segmentation}  we present the   spectral clustering methodology   and   the formalism to implement it  on $S^2$. In Sect.~\ref{sec:spectral}, we present our implementation of spectral clustering that makes use of the HEALPix sky tesselation. In Sect.~\ref{sec:results}  we show two applications of  how the patches defined with spectral clustering can be used to improve the performance of parametric component separation techniques. We finally discuss results and cosmological implications in  Sect.~\ref{sec:discussion} and \ref{sec:conclusions}.

\section{Spectral image segmentation}
\label{sec:segmentation}

Image segmentation via spectral clustering is an  efficient technique that  combines spatial similarity (or affinity)  of pixels with   the similarity based on image-related characteristics of the pixels \citep{Zhang2018,Zelnik-Manor}. There are profound geometric and analytical reasons for this  and are concisely discussed  in the dedicated Appendix \ref{appendixA}.

Any image can be mathematically encoded by means of a real-valued (ideally smooth) function $f$, which is defined on a portion of a surface and  sampled with a fixed resolution given by the pixel grid. For example a grey-scale image is  encoded by one function, whereas RGB images are encoded by three. 

We summarize below the standard steps for a spectral clustering process:\medbreak
\begin{enumerate}
    \item  Translate an image into a graph by defining  a suitable \emph{adjacency} condition. A typical choice is that any pixel is connected to its nearest-neighbours. For example, in correspondence to pixel $j$ neighbour  of  $i$, we  can set to a non-zero value the $ij$ element  of the so called \emph{adjacency} matrix $A_{ij}$.  Once the connections among all pixels are defined via the weights on the adjacency matrix we can build a graph  by choosing a conventional order of the vertices (pixels) $x_i$. The way  the adjacency weights are assigned  is usually based on accounting for  the geometric distance or the  distance between the values that $f$  takes in each pixel. An example of weights estimated from a distance is commonly weighting the graph nodes  by a Gaussian function of the distance between two  pixels.

\item  Compute the graph Laplacian matrix from the weighted adjacency matrix. We use the symmetric random walk Laplacian:
\begin{equation}
   L_{sym} \equiv I-D^{-\frac{1}{2}}A D^{-\frac{1}{2}}   
   \label{eq:laplacian_sym}
\end{equation} \noindent where the $I$ is the identity matrix and $D_i=\sum_j A_{ij}$ is a diagonal matrix called \emph{degree} matrix.
\item Compute the eigenvectors of the graph Laplacian, $L_{sym}$ and select a subset of $n$ eigenvectors,  in  correspondence to the ones that contribute the most to the  eigenspectrum of $L_{sym}$.  
E.g. for the specific case of 2D images, each eigenvector can be  visualized as images encoding large (small) spatial variations depending whether the correspondent eigenvalue is small (large).  
 This process defines an \emph{embedding} of the graph associated to the image in $\mathbb{R}^n$, i.e.  each pixel $i$ in the image is characterized by $n$ {features}, i.e. the value of  each  eigenvector    in the $i$-th  pixel.
\item  
Once the embedding in $\mathbb{R}^n$  is defined,   the standard Euclidean distance  in $\mathbb{R}^n$  is evaluated for all the pair of pixels in order to  run a suitable \emph{Agglomerative} or \emph{Divisive}  clustering algorithm. 
\end{enumerate}

In this work, we combine the spectral clustering and adapted it in the context of quantities defined  in the $S^2$ manifold.   This is  generally  referred to as \emph{manifold learning}:  a discipline that combines statistical methods with techniques developed in differential geometry. It is based on the assumption that point clouds of multivariate $n-$dimensional variables are sampled on or close to smooth compact sub-manifolds of $\mathbb{R}^n$ (  the existence of a boundary is usually ignored). In Appendix~\ref{appendixA}, we outline more in detail how manifold learning can be combined with  spectral clustering. However, we remark here  two relevant and non-standard aspects of the implementation of the method proposed in this  paper that make the geometric nature of spectral clustering very relevant:
\begin{itemize}
\item  Maps from very wide astronomical surveys  are sampled on the sphere, we therefore implement the   affinity between pixels  accounting for the angular distance on the  $S^2$ sphere (presented in Subsect.~\ref{subsec:heat_healpix}). 
\item  We choose the partion as the one that minimizes the within- and between-cluster variances. We present in Subsect.~\ref{sec:clustering},  how the variance depends on several partition choices and eigenvector bands. 
\end{itemize}

In order to better understand the details in the following sections, we briefly summarize few propositions in the context of Graph Theory which will be used   to justify  the choice for the adjacency in Sect.~\ref{sec:spectral}. 
 
\begin{Proposition}
The construction of the  graph Laplacian matrix $L$ is  known to   approximate  the Laplace-Beltrami operator $\Delta_g$ on a Riemannian manifold embedded in $\mathbb{R}^n$ and sampled in a grid of points (e.g. our set of pixels).  Therefore, the weighted adjacency matrices can be directly derived from the   integral kernel (or Green function)  of the Laplacian operator in the sense of Fredholm's theory \citep{fredholm}. 
\label{prop:1}
\end{Proposition}

\begin{Proposition}
 The Laplace-Beltrami operator on a compact differentiable manifold $M$ has discrete spectrum $\lambda_0\leq \lambda_1\leq\lambda_2\leq...\leq\lambda_n $ and its eigenfunctions $\varphi_i$ form an orthonormal basis of the space $L^2(M)$. 
Analogously, every discrete approximation of a function on a manifold can be expressed with respect to a basis of  eigenvectors of the graph Laplacian. \label{prop:2}
\end{Proposition} 
 
As a consequence of Proposition~\ref{prop:1} and \ref{prop:2}, estimating the  eigenspectrum  of suitable graph Laplacian matrices sampled in a grid  approximates the spectrum  (eigenvalues and eigenfunctions) of  the Laplace-Beltrami operator.

 Moreover, in defining the optimal image segmentation it is quite common   to define  the \emph{ Dirichlet energy functional}, i.e. 
\[ E[f]\equiv\frac{1}{2} \int_M \parallel \nabla f (x)\parallel ^2 d\omega_g\]
that measures the spatial variability (or  \emph{spatial frequency})  of a smooth function on a manifold $M$.  Thus, since  the Dirichlet energy associated  to   the eigenfunctions of the Laplace-Beltrami operator is $E(\varphi_i)=\frac{1}{2}\lambda_i$ with monotonically increasing Dirichlet energy\footnote{A common example of increasing Dirichlet energy eigenfunctions can be found among the  eigenfunctions of the Laplace-Beltrami operator in the sphere, i.e. the Spherical Harmonics, $Y_{\ell m}$.  We observe  increasingly larger spatial variability of the $Y_{\ell m}$  in correspondence to larger values of multipole number $\ell$, with  fluctuations involving smaller and smaller angular scales.} (Proposition~\ref{prop:2}), the set of the first $n$ eigenfunctions provides an  optimal embedding of $M$ in $\mathbb{R}^n$   corresponding to the  minimum  Dirichlet energy\footnote{Similarly, this properties propagates to the eigenvectors of the Laplacian matrix (eq. \eqref{eq:laplacian_sym}).}. 
Generally, this process can be  also addressed by selecting a subset of eigenfunctions (corresponding to a certain  \emph{ Dirichlet energy band}) to embed $M$ in a $n-$dimensional space.

The choice of the  eigenfunctions  related to a specific  Dirichlet energy band  can impact  the quality of the image segmentation. This is mainly due to the fact that the clustering methodology is based on the choice of the eigenvectors of the Laplacian matrix, to which   corresponds a certain degree of spatial variability, or \emph{granularity}.   We devote Sect.~\ref{sec:clustering} to describe how this applies to the case presented  in this work and how the choice of the energy  band affects the characteristic size of the clusters.

The key ingredient in this framework is the definition  of the  adjacency, from which the Laplacian matrix can be then derived via eq.~ \eqref{eq:laplacian_sym}.   %
In the next section, we describe extensively the adjacency adopted in this work.

\section{Spectral Clustering on HEALPIX maps} \label{sec:spectral}
In this section, we present an implementation of spectral clustering applied on images defined on  the full celestial sphere. We adopt maps following the \textsc{HEALPix}  scheme \citep{2005ApJ...622..759G,Zonca2019}.

\subsection{Choice of  the adjacency} \label{subsec:heat_healpix}

Let us denote by $X=(\mathbf{x}_1,\dots, \mathbf{x}_{n_{\rm pix} } ) $, a set of  normed-1  vectors (with $\mathbf{x}_i \in \mathbb{R}^3$), encoding the coordinates of each \textsc{HEALPix} pixel.
Given $X$, we can construct a matrix  as $  \Theta  = X^T  X $,  being a $n_{\rm pix} \times n_{\rm pix} $ matrix, with diagonal equal to $1$, and off-diagonal elements  encoding the scalar product estimated pairwise on the columns of $X$. 
Notice that $\Theta$ encodes the cosines of the scalar product as the columns of $X$ are normed-1. 

We can  therefore define the degree of affinity  between  pixels in a \textsc{HEALPix} map from $\Theta$, as: 
\begin{equation}
    A \propto   \rm{exp}\left({-(1- \Theta)^2 /2\sigma_{\rm pix}^2}\right),
\label{eq:adj_heat}
\end{equation}
where $\sigma_{\rm pix}$\footnote{In this work, we adopted   \texttt{nside=32} which results in $\sigma_{\rm pix}\sim 46'$.}  is related to the pixel scale of the map.   	By expanding eq.~\eqref{eq:adj_heat} in the   Harmonic domain in terms of the Legendre Polynomials $\mathcal{P}_{\ell}$, as
\begin{equation}
    A \equiv\sum_{\ell =0} ^{+\infty } \frac{2\ell +1}{4\pi}   e^{-\ell(\ell+1)\sigma_{\rm pix}^2} \mathcal{P}_{\ell}( \Theta),
\label{eq:adj_heat_harm}
\end{equation}
 we observe that the scale given by $\sigma_{\rm pix}^2$  sets a threshold  $\ell_{\rm max} $  to which we can stop the sum in eq.~\eqref{eq:adj_heat_harm}.

\begin{figure} 
    \centering
    \includegraphics[scale=.4, trim=0 .0cm 0 0 ,clip=true ]{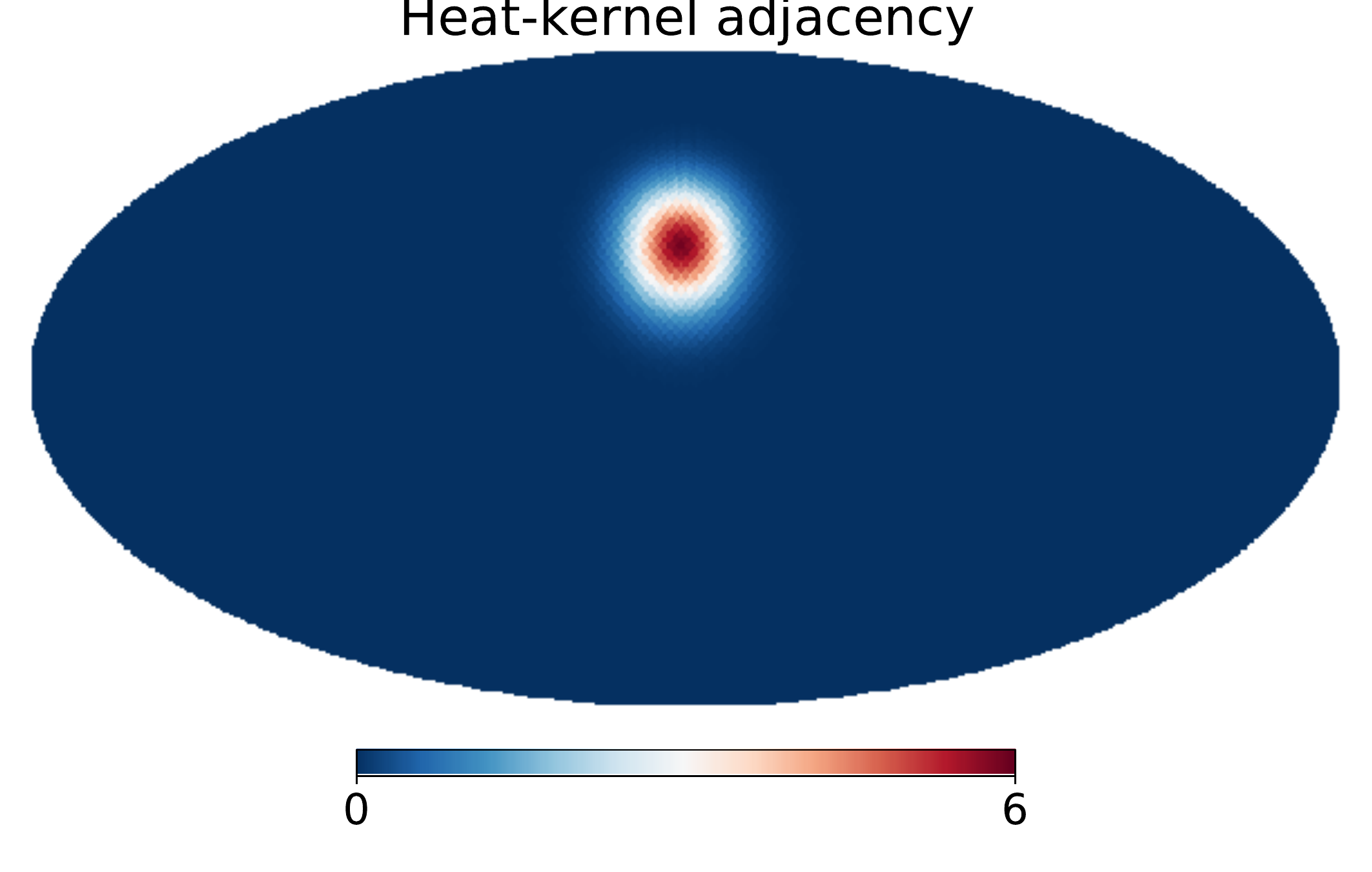}
    \caption{Adjacency as estimated in eq.~\eqref{eq:heat_kernel} relative to the pixel centered at $(\ell, b)= (0,30)$ deg Galactic coordinates. We recall the reader that $A_{ij}$ is a $n_{\rm pix} \times n_{\rm pix} $, so its rows and/or columns can be visualized as \textsc{HEALPix} maps.}
    \label{fig:kernel}
\end{figure}

In Appendix~\ref{app:heat_kernel}, we show how the functional form in eq.~\eqref{eq:adj_heat_harm} derives from the integral kernel (also known as \emph{heat kernel}) of the Laplace-Beltrami operator, $\Delta _g$,   in $S^2$ \citet{zhao2018}.
We show in Fig.~\ref{fig:kernel} a row of the affinity matrix $ A$ as defined in eq.~\eqref{eq:adj_heat}. 

\begin{figure} 
    \centering
    \includegraphics[scale=.65, trim=.7cm .6cm 0 .6cm, clip=true ]{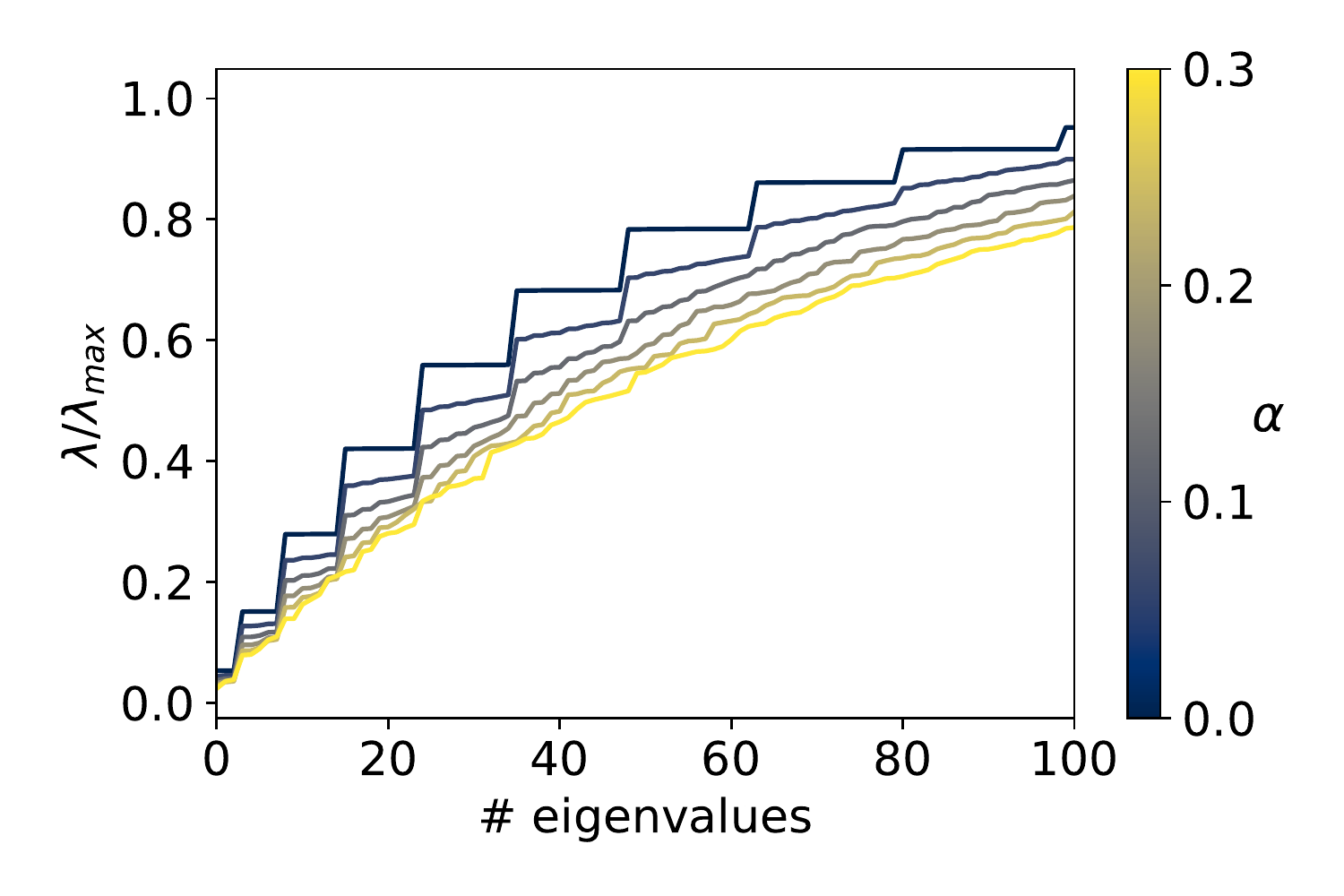}
    \caption{Eigenvalues of Laplacian matrix estimated  as   in eq.~\eqref{eq:laplacian_sym}, for several choices of the scaling factor $\alpha$.}
    \label{fig:evals_deform}
\end{figure}

 \begin{figure*} 
    
     \includegraphics[scale=.38, trim=0cm 1.cm 0 .7cm, clip=true ]{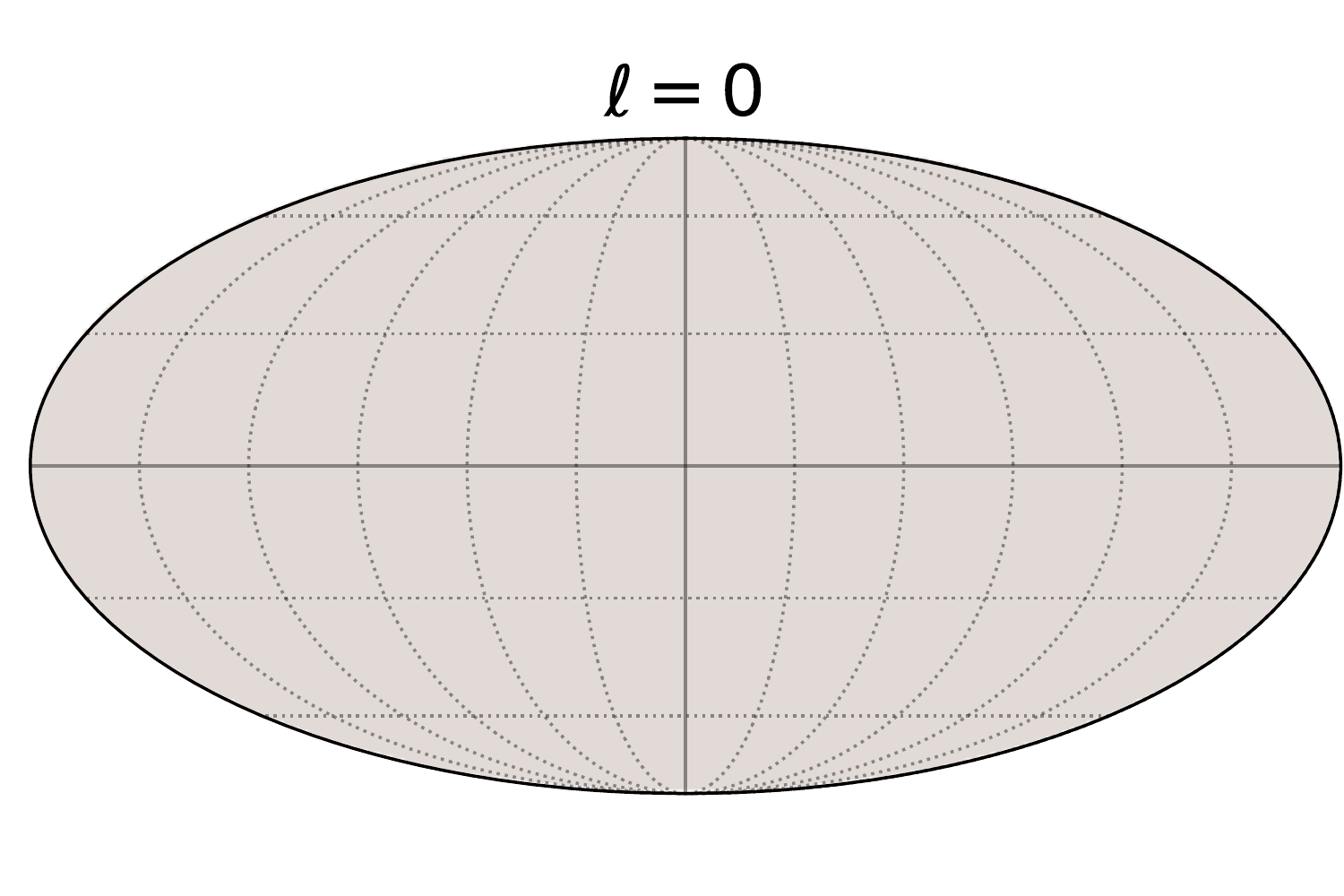}
      
    \includegraphics[scale=.47, trim=0 2cm 0 2.2cm, clip=true]{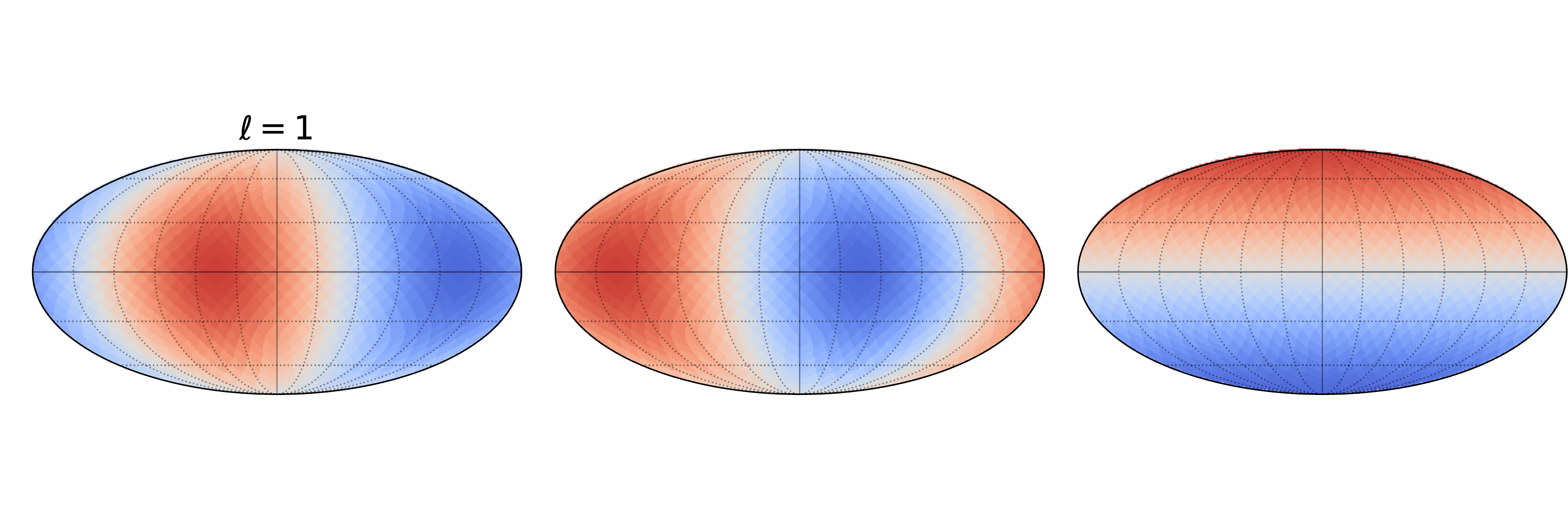}
    \includegraphics[scale=.47,trim=0 4cm 0 4cm, clip=true]{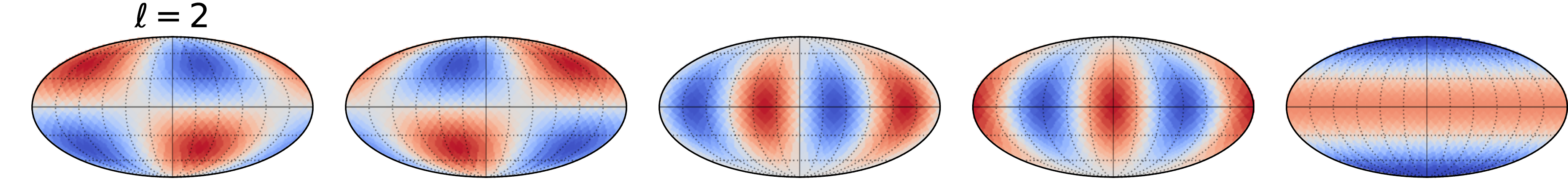}
    \includegraphics[scale=.47
    ,trim=0 4.7cm 0 4cm, clip=true]{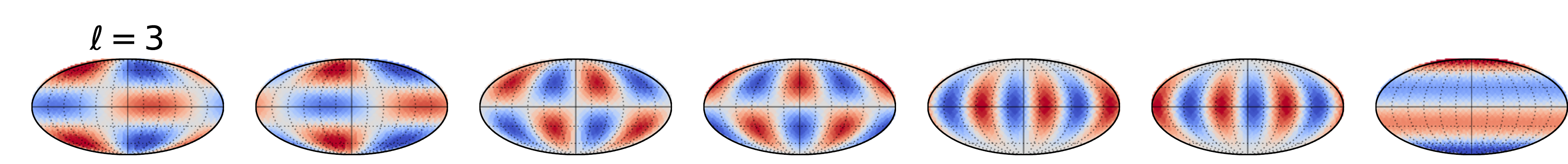}
     \caption{The first eigenvectors of $L$  corresponding to $\ell=0,\dots,3$ and assuming no uncertainty deformation (i.e. $\alpha=0$). The colorbar ranges for all the maps between $\pm 0.07$. }
     \label{fig:evec_heat}
 \end{figure*}
 
 \begin{figure*} 
    
     \includegraphics[scale=.38, trim=0cm 1.cm 0 .7cm, clip=true ]{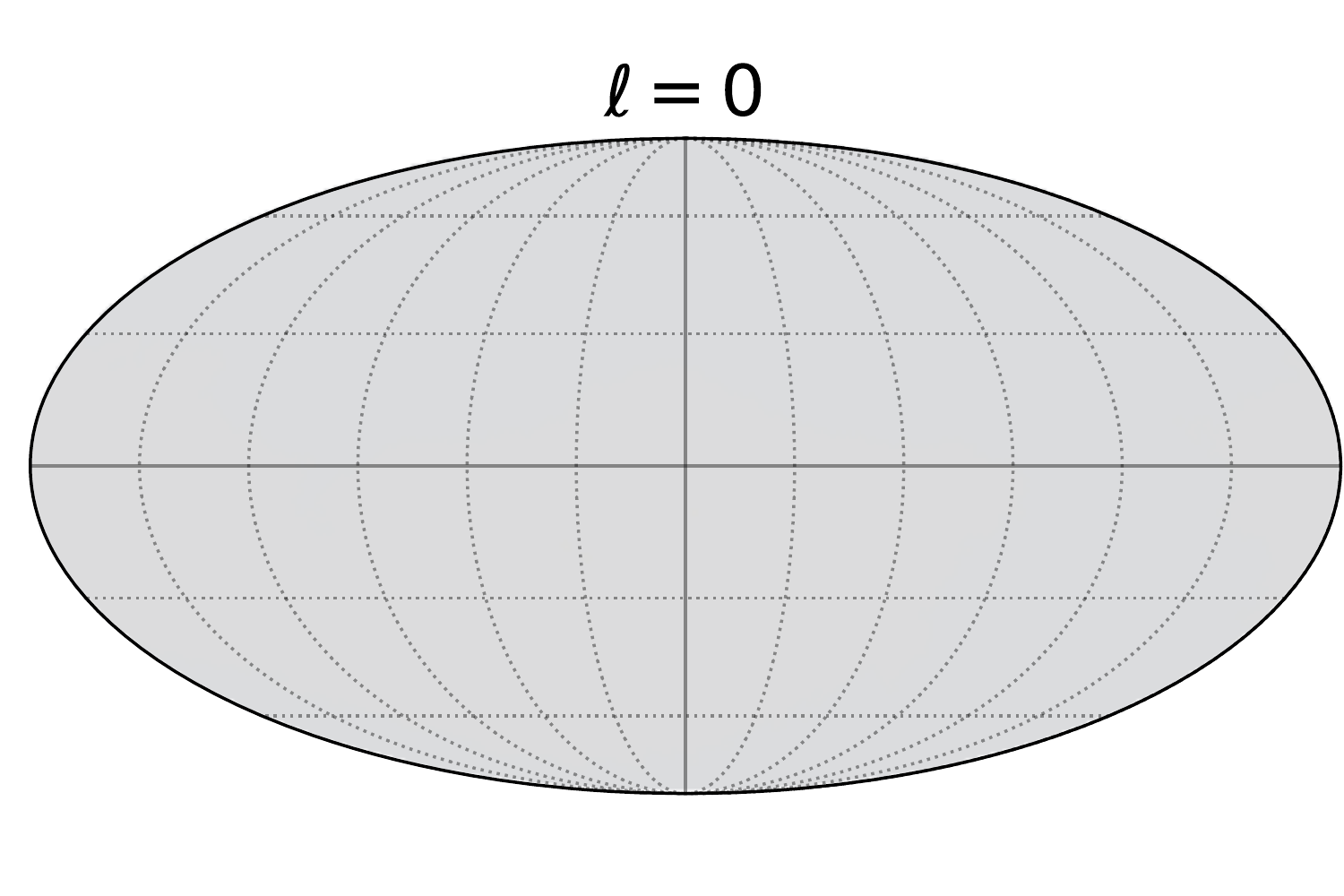}
      
    \includegraphics[scale=.47, trim=0 2cm 0 2.2cm, clip=true]{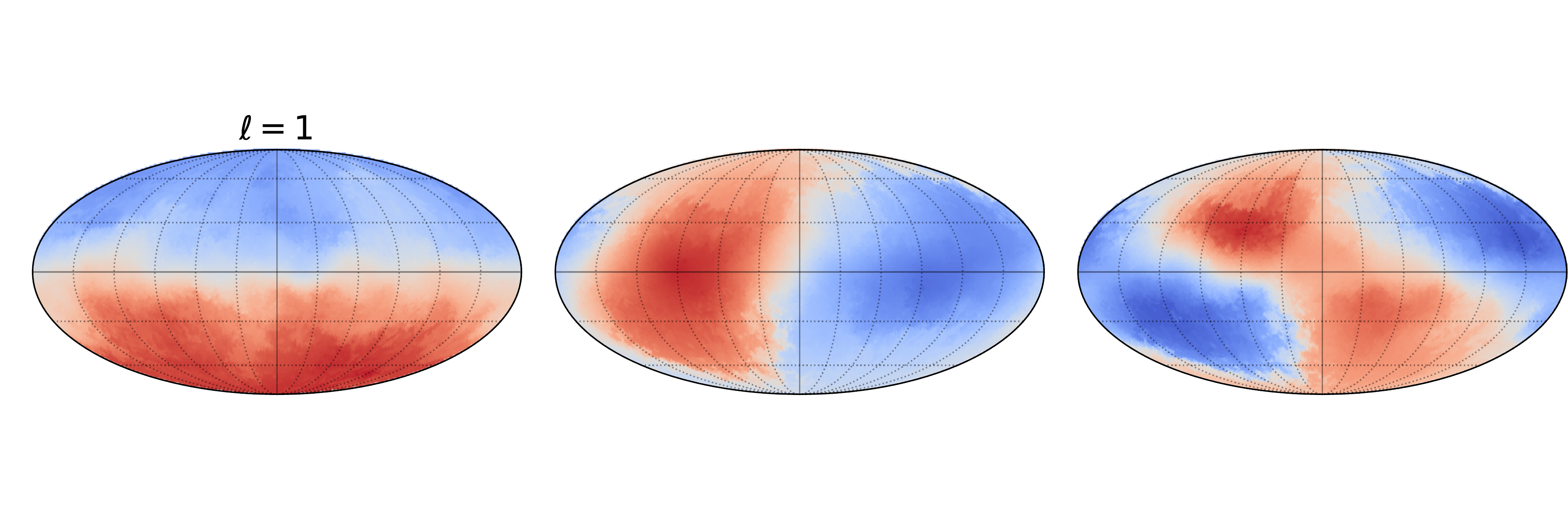}
    \includegraphics[scale=.47,trim=0 4cm 0 4cm, clip=true]{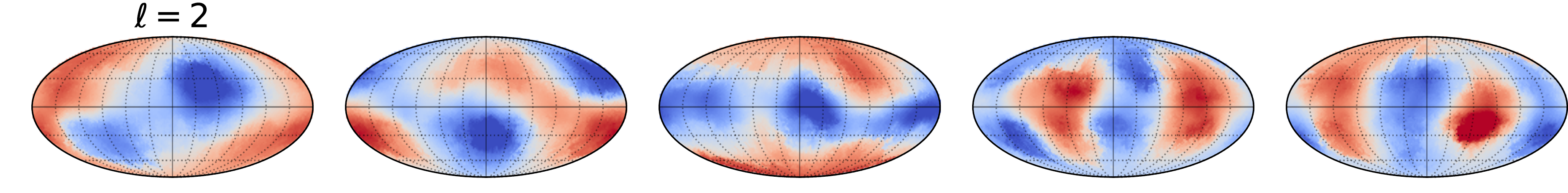}
    \includegraphics[scale=.47
    ,trim=0 4.7cm 0 4cm, clip=true]{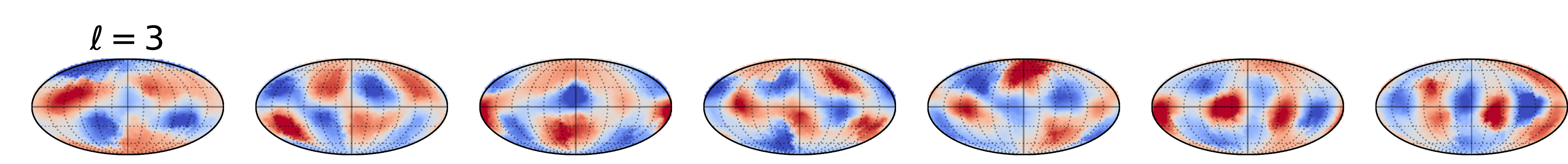}
     \caption{The first eigenvectors of $L$  with  affinity  deformation $\alpha=0.25$.  For sake of comparison with Fig.~\ref{fig:evec_heat} we report also in this plot the   correspondence to $\ell=0,\dots,3$ although the analogy lose specific meaning for $\alpha \neq 0$.   
     The colorbar ranges for all the maps between $\pm 0.02$. 
     }
     \label{fig:evec_heat_def}
 \end{figure*}

When clustering methodologies are employed, it is  common to   ask oneself how the  adjacency  in eq.~\eqref{eq:adj_heat} is affected  in presence of  the uncertainties in the measurements at a given pixel location. 
 E.g. suppose that we have measurements of the \emph{modified black-body}   spectral index of dust emission, $\beta_d$, and its uncertainties, $\sigma_d \equiv \sigma (\beta_d)$.
 We require that two pixels encoding statistically compatible values of $\beta_d$ would be more easily associated together into a single cluster  with respect to  pixels with incompatible values.  
 
We therefore want to identify a way to combine the measurements and the uncertainties  into the  heat-kernel adjacency (eq.~\ref{eq:adj_heat})  given realistic  measurements  of  foreground spectral parameters. For the following application,  we will use the map of the dust spectral index $\beta_d$ and the uncertainty map $\sigma_d $  obtained  from the \planck data processed  with the COMMANDER separation algorithm \citet{pldiffuse2015}.

Intuitively,  we can distort   $\Theta$ defined in eq.~\eqref{eq:adj_heat} with  a certain  weight given  by  the measurements  and the uncertainty of the parameter  at a given pixel location.
 Given the state of art data on Galactic  foregrounds in the sub-mm regime,   we consider  the measurements in each pixel $i$ to be  normally distributed around $\beta_{d,i}$   with a width given by its uncertainties,  $\sigma_{d,i}$  \citep{pldiffuse2015}. We therefore  generate a mock sample of  100 Gaussian random  numbers given the features $(\beta_{d,p},\sigma_{d,p})$, and we estimate the two-sample Kolmogorov-Smirnov (KS) test onto  pairs of different pixels $i$ and $j$. This allows us  to test the null hypothesis that  the distribution of $\beta_d^i$ values is drawn from the same underlying distribution as those of $\beta_d^j$ within an assumed confidence level. We repeat the test $100$ times, evaluate the statistical quantile $Q_{ij}$ of each  KS test, and set the median value all the KS tests, $Q_{ij}^{med}$  as the adjacency weight evaluated for the $ij$ pixel pair. Thus, if $Q_{ij}^{med}$ is   small enough to  reject the null hypothesis at a $2\sigma$ confidence level, the connection is  weighted with a  very  low   weight and it is unlikely that  the two  pixels will be associated into the same cluster.  On the other hand, when   $Q_{ij}^{med}\sim 1  $ we cannot reject the null hypothesis and the two pixels can be associated, as they are \emph{ parallel} in this metric.
 
 In order to combine the adjacency from the KS test and the one defined in eq.~\eqref{eq:adj_heat}, we can  treat the KS quantiles as cosine of angles, so that  a straightforward deformation of $\Theta$ can be:  
 \begin{equation}
     \Theta'   = \cos \left( \arccos (\Theta   ) + \alpha \left(1 - Q^{med}\right)\frac{\pi}{2} \right), 
     \label{eq:deform}
 \end{equation}
 where $0<\alpha\leq 0.5$ is a \emph{scaling factor} parameter that  weighs the relative contribution of the KS quantile similarity  with respect to the geometrical one (the proximity of two pixels on the sky).{ Values of $\alpha>0.5$ result into too large distortions that   break  the metric properties of eq.~\eqref{eq:deform} and make   the Laplacian  a singular matrix (see  next Subsect.~\ref{sec:clustering}).} The   updated cosine matrix $\Theta'  $ is then finally inserted into eq.~\eqref{eq:adj_heat} to evaluate the adjacency weights obtained from this distortion. 
 
 Given the definition of adjacency (eq.~\eqref{eq:adj_heat} or ~\eqref{eq:deform}),  we can therefore estimate the Laplacian matrix as in eq.~\eqref{eq:laplacian_sym} and estimate its   first $N_{eigen}$ eigenpairs, related to the smallest eigenvalues. 
We remark that we do not factorize the whole Laplacian matrix, since we are  interested in a subset of  eigenvectors.  We thus  approximate the  eigenpairs by means of the so called  \emph{ Ritz approximation}, a technique  which has already been exploited  in previous  literature  \citep{szydlarski_2014,puglisi_2018} to approximate very well  the exact eigenpairs of a matrix.

 In Fig.~\ref{fig:evals_deform}, we show the first 100 Ritz eigenvalues $\lambda$ of $L$ for different choices of $\alpha $. 
 We firstly focus on the $\alpha=0$ case. 
As already mentioned above,  the functional form of the  adjacency in eq.~\eqref{eq:adj_heat} descends from  the integral kernel of the Laplacian operator in $S^2$,  which specifically coincides  to the \emph{angular momentum } operator $\hat{L}^2$ in quantum mechanics (see Appendix~\ref{app:heat_kernel} for further details).   We thus expect  the eigenvectors to be exactly the eigenfunctions of $\hat{L}^2$, i.e.  the Spherical Harmonics and the multiplicity of the eigenvalues to respect the same algebraic multiplicity: for a given multipole number $\ell$,  its multiplicity goes as $2\ell +1$, in correspondence of the $m$ azimuthal  multipole number.   
 
 \noindent In fact,  we    notice that for the $\alpha=0$ case (solid blue) line, the   algebraic multiplicity  grows exactly   with   $2 \ell +1$. 
 Furthermore, the eigenvectors shown in Fig.~\ref{fig:evec_heat} perfectly resemble the  spherical harmonics. Both the eigenvalue degeneracy as well as the morphology of the Laplacian eigenvectors represent a remarkable  validation test for the overall implementation described above.   

On the other hand, the distortion introduced by $\alpha \neq 0$ breaks  the Laplacian eigenfunction algebraic multiplicity, so that the eigenspectrum becomes  gradually strictly   monotonic   for higher values of $\alpha$.  As a consequence,  the eigenvectors as well deviate from the spherical harmonics, being more and more weighted by the measurement uncertainties. In our specific case involving  Galactic emission, we indeed observe several anisotropies corresponding to the  Galactic plane as shown  in Fig.~\ref{fig:evec_heat_def} for the  case $\alpha=0.25$.

  \begin{algorithm}
\caption{Spectral Clustering Optimization of a field $X$ with uncertainty $\sigma(X)$}
\label{alg:spectral_clus}
\begin{algorithmic}[1]

\Procedure{Spectral Clustering }{$X,\sigma(X) $}       
    \State  Parameter Affinity, $Q$ Initialization with ($X,\sigma(X) $) data  
    \For{$\alpha$  in $[0,\dots,\, \alpha_{max}]$} 
    \State Build  Affinity matrix  with $\alpha$ distortion (eq.~\eqref{eq:deform} )
    \State Build Laplacian matrix, $L$ (eq.~\eqref{eq:laplacian_sym})
    \State Estimate Ritz eigenpairs 
    \State Select an eigenvector-band, $U$
    \State Estimate the  euclidean affinity $E$ from the columns of $U$
    \For{$\delta $ in $[0,\dots, \delta_{max}]$ } 
        \State Run Agglomerative clustering 
        \State Estimate $W(\alpha, \delta)$  
        \EndFor
    \EndFor
    \State Find local minima of $W$, $(\alpha_*, \delta_*)$
    
\EndProcedure

\end{algorithmic}
\end{algorithm}

\subsection{Identifying the optimal partition}\label{sec:clustering}
  
  In this subsection, we aim at discussing how the clusters are estimated and how to identify the range of partition optimality.  
  The overall spectral clustering algorithm described  in the   sections above is summarized  in  Algorithm~\ref{alg:spectral_clus}. In particular, we employ the implementation of  agglomerative    clustering publicly available in the \textsc{Scikit-Learn} python package\footnote{\url{https://scikit-learn.org/stable/modules/generated/sklearn.cluster.AgglomerativeClustering.html}} to estimate the clusters. 
 
\begin{figure*}
    \centering
    \includegraphics[scale=.29, trim=0cm 0cm 5.7cm 0cm, clip=true]{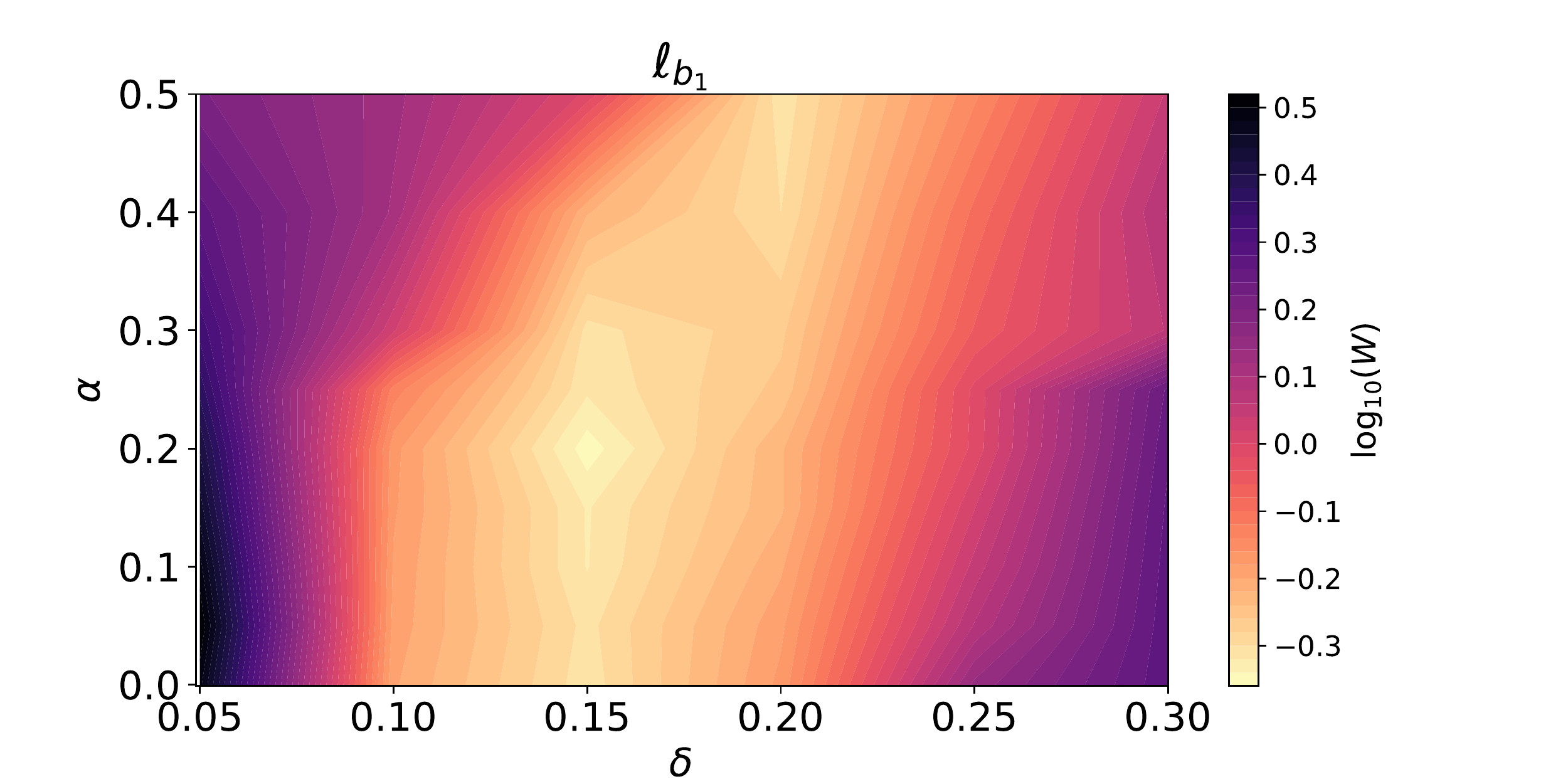}  
     \includegraphics[scale=.29, trim=1.9cm 0cm 5.7cm 0cm, clip=true]{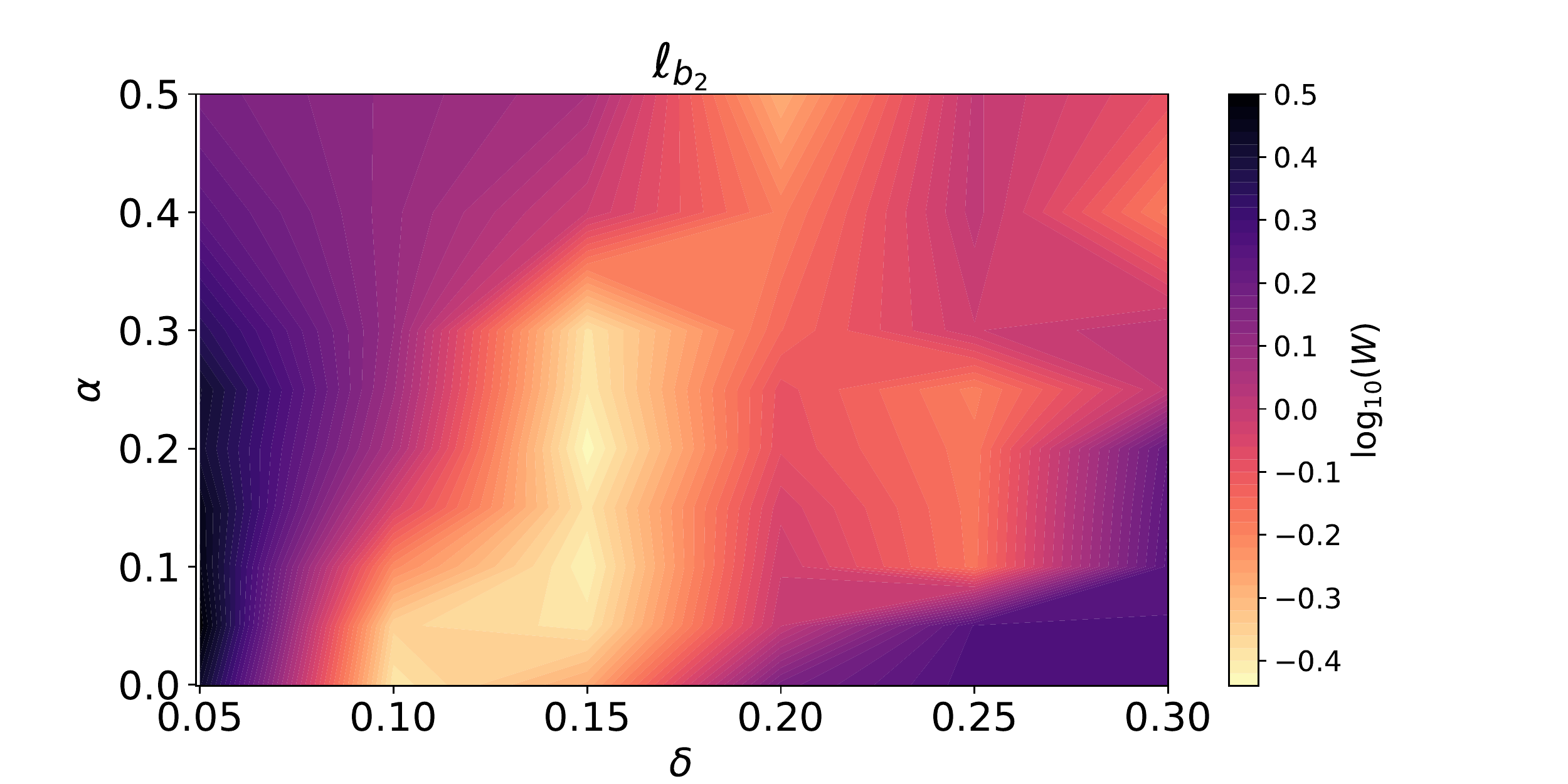}  
      \includegraphics[scale=.29, trim=1.9cm 0cm 0cm 0cm, clip=true]{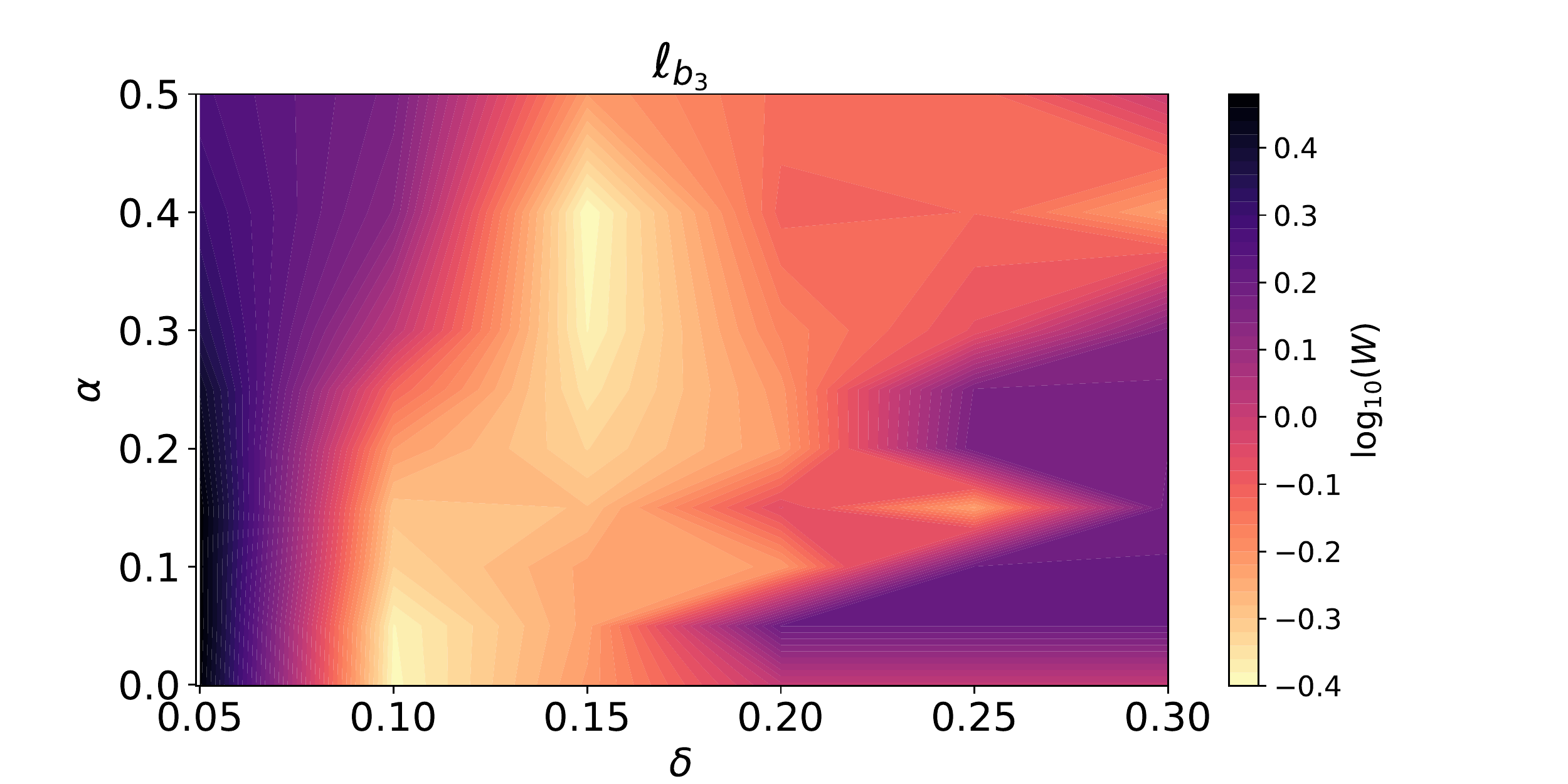}  
    \caption{Variance $W$ as in eq.~\eqref{eq:variance} in the $\alpha-\delta$ space, evaluated  in  clusters constructed with the   3 eigenvector bands as defined in the main text. For the analysis shown in Sect.~\ref{sec:results},  we adopted the $\ell_{b3}$ being the band that  accounts for the eigenvectors encoding most of the angular  scales to be clustered. }
    \label{fig:clus_variance}
\end{figure*}

  As shown in Fig.~\ref{fig:evec_heat},  the first  eigenvalues  correspond to eigenvectors with longer oscillation in the map, and vice versa. Furthermore, we observe that the spectrum tends to saturate to a maximum value $\lambda_{max}$  after  about $\sim 100 $ values (see Fig.~\ref{fig:evals_deform}). Since we want to identify the smallest eigenvalues encoding  most  of  the bulk of the Laplacian eigenspectrum, we  consider    the first  $n_{eig} = 256$  Ritz eigenpairs and therefore build the matrix  $U \in \mathbb{R}^{n_{pix} \times n_{eig}}$ containing  $n_{eig}$ eigenvectors as columns. 
 
 {We remark here that the eigenvector  matrix $U$ is critical in the context of spectral clustering because it provides the embedding of $S^2$ into  $\mathbb{R}^{n_{eig}}$ space spanned by the Laplacian eigenvectors.
Moreover,  it helps in reducing  the dimensionality of our problem from $n_{pix}$ features, i.e. the number of features we have in a map, to $n_{eig}$ and  in avoiding the  \emph{curse of high dimensionality} which clustering methodologies might frequently incur}. 
 
 We then consider each row of $U$, $(u_{i} )_{i=1,\dots, n_{pix} } $  as our data vector and we perform agglomerative clustering  with    $n_{pix}$ samples and  $n_{eig}$ features. 
 We measure the pairwise similarity, $E_{ij}$,  between two vectors $u_i$ and $u_j$ (representing the $i$-th  and $j$-th pixels)  by estimating   the euclidean distance between them.  The two pixels are  associated to a group if $E_{ij}<\delta$, with $\delta$ being a free parameter, commonly referred as  the \emph{linkage distance threshold}. 
 Small values of  $\delta$ tend to typically produce small clusters, and vice versa. This can be intuitively understood as follows: the more constraining the distance  threshold, the  harder it is to associate two pixels  together into a cluster. 
 
  We notice that for several values of $\alpha$ the  distance matrix range changes in such a way that we get   smaller eigenvector similarities for larger values of $\alpha$.
  This is a consequence of the fact that the deformation of adjacency with KS weights (eq.~\ref{eq:deform})  results into a compression of the Laplacian eigenspectrum to  smaller  condition numbers.  
In fact, we notice a trend    in the Laplacian eigenspectrum to i) lose the typical $2\ell +1$ degeneracy for $\alpha\neq 0$,  ii) flatten to smaller condition number, iii)  to become degenerate to a small  value numerically close to zero.  As a results the convergence of the Laplacian Ritz eigenpairs estimation time increases for $\alpha >0.3$. \footnote{This is the reason why we  set the maximum value of  $\alpha$ to $0.3$.}

Furthermore, the eigenvector distance  for several  values of $\alpha$ needs to be normalized   in such a way that we use 
 use the same range in $\delta $ to compare the variances estimated from each run of  the agglomerative clustering.
   We therefore rescale the  eigenvector distance $E_{\alpha}$  evaluated  for different choices of $\alpha$ as it  follows: 
  \begin{equation}
      E'_{  \alpha} =  \frac{ E_{  \alpha}/M   - m} {1-m }, 
      \label{eq:rescaling}
   \end{equation}
  with $M= E_{{\rm max}, \alpha}$    and $m=E_{{\rm min}, \alpha}$. We note that  $E'$ ranges from $
      E'_{\rm max}=   1$ to $   E'_{\rm min}=  C m$, 
 with $C=(1- M) / M \sim {\rm constant}$ for different values of $\alpha$.  The rescaling in eq.~\eqref{eq:rescaling} can be seen as a sort of \emph{rigid}  scaling of the distance, making the  distance matrices for different choices of $\delta$ and $\alpha$ to be comparable without loosing too much information.  { Notice that  eq.~\eqref{eq:rescaling}   scales the maximum of the distance to 1  but we make sure that it does not affect the shape of distance distribution. We also make sure that  the   ratio of minimum distances for  different choices of $\alpha$ is the same as before the rescaling in eq.~\eqref{eq:rescaling}.}

 Furthermore, we  consider 3 bands to identify the eigenvectors with major contribution to the total   Dirichlet energy budget, i.e. the spatial variability, in the domain partition:

\begin{itemize}
    \item $\ell _{b1}$, including the first 100 eigenvectors, corresponding to $1\lesssim \ell\lesssim 10  $, 
    \item $\ell _{b2}$, including vectors  between the 90-th and 150-th eigenvectors, corresponding to $9\lesssim \ell\lesssim 12  $,
    \item $\ell _{b3}$, sampling the whole eigenspectrum every other 5 vectors  up to the 175-th  one, corresponding to $1\lesssim \ell\lesssim 14 $.
\end{itemize}

 \noindent We create a grid with different points evaluated in  the $\alpha-\delta $ parameter space, with $\alpha \in [0, 0.5]$, $\delta\in [0.05,0.3 ] $,  and estimate the cluster variance, $W$, defined as
 \begin{equation}
     W \equiv \sqrt{V_w^2 + V_b^2},
     \label{eq:variance}
 \end{equation}
  with $V_w$  being the within-cluster variance,
 \begin{equation*}
     V_{w} \equiv \sum _{k =1} ^{K } \sum_{u_k \in C_k} (u_k - \mu_k )^2, 
 \end{equation*}
  $V_b$ the between cluster variance,
  \begin{equation*}
     V_{b} \equiv \sum _{k \neq k' }  (\mu_{k'}  - \mu_k )^2,
 \end{equation*}
 $K$  the total number of clusters and  $\mu_k$ the   centroid of cluster $C_{k}$.

Fig.~\ref{fig:clus_variance} shows $W$ defined in $\alpha-\delta $ space evaluated for the 3 different eigenvector bands. The optimal  parameters  $(\alpha_*, \delta_*)$ correspond to local  minima of $W$. Notice that the range of optimality changes for different    eigenvector bands as different scales are encoded in each band. We emphasize that similar optimal ranges are found among the three bands, indicating that the clustering is stable with respect to the  choice of the band.   

Since  the $\ell_{b3}$ band encodes scales from  most of the Laplacian eigenspectrum, we adopt this band  as a baseline for the applications  presented in the following sections. 

Finally, we notice that the   optimality,   shown in Fig.~\ref{fig:clus_variance}, does not correspond to a very localized region in the parameter space spanned by $\alpha$ and $\delta$. Rather it shows as a narrow vertical band at around a certain value of $\delta$ for which several choices of $\alpha$ can be equally optimal. We find that intermediate values of $\alpha\sim 0.15$  lead to a balanced trade-off between defining the clustering given the features of spectral parameters   and the intrinsic adjacency in $S^2$. We refer the reader to Appendix~\ref{app:optimality} for  further details.  

\vspace{0.5cm}

The estimation of the Laplacian adjacency and eigenpairs and the optimization of   spectral clustering  has been performed  on 300 Cori KNL nodes of the NERSC Supercomputing facility\footnote{\url{https://www.nersc.gov/}}. It takes   $\sim4000$ cpu-hours to execute the overall procedure outlined in Algorithm~\ref{alg:spectral_clus} with maps at \texttt{nside=32}.

\section{Applications}\label{sec:results}
    
We present two applications of our spectral clustering algorithm for the purpose of performing parametric component separation. First, in Sect.\ref{subsec:compsep}, we apply the spectral clustering algorithm to derive a partition of the sky from  publicly available maps of synchrotron  and dust spectral parameters. Once the patches are defined we  perform a parametric component separation on  each patch and assess  the   post-foreground cleaning  residuals in the CMB $B-$mode  maps. Secondly, in Sect.~\ref{sec:nc}, we 
identify the patches by means of an ancillary dataset: the number of HI clouds along the line of sight, $\mathcal{N}_c$, presented in \citet{Panopoulou2020}. This dataset is an independent tracer of the foreground dust emission. We then perform  component separation within these patches and estimate the quality of the reconstruction.

\subsection{Clustering applied on Galactic foreground spectral parameters }\label{subsec:compsep}

\begin{table*} 
\caption{Nominal specifics of   CMB  experiments to run component separation with \textsc{FGBuster}.   }\label{tab:specs}
	\begin{tabular}{lccc}  
		\hline
 & Frequency [GHz] & Sensitivity  $\left[ \mu\mathrm{ K \, arcmin}\right]$  & FWHM $[{\rm arcmin}] $  \\
\hline 
\multirow{2}{*}{\sosat} & $27, 39, 93$  & $35,21,2.6  $  & $ 91,63,30 $   \\
                        & $145, 220, 270$ & $3.3,6.3,16  $  & $ 17,11,9$  \\
&&& \\
\multirow{3}{*}{\lb } & $40,   50,   60,   68,   78$ & $ 37.5,24,19.9,16.2,13.5,  $  & $69,56,48,43,39$  \\ 
& $89, 100,119, 140,166$ & $ 11.7,  9.2,7.6,5.9,6.5 $& $35,29,25,23,21 $  \\
& $195, 235,280,337,402$ & $5.8,7.7,13.2,19.5,37.5$& $20,19,24,20,17   $  \\
\hline 
\end{tabular}
\end{table*}
 
\begin{figure*}
    \centering
    \includegraphics[scale=.5, trim =2cm .3cm 0 .4cm,clip=true] {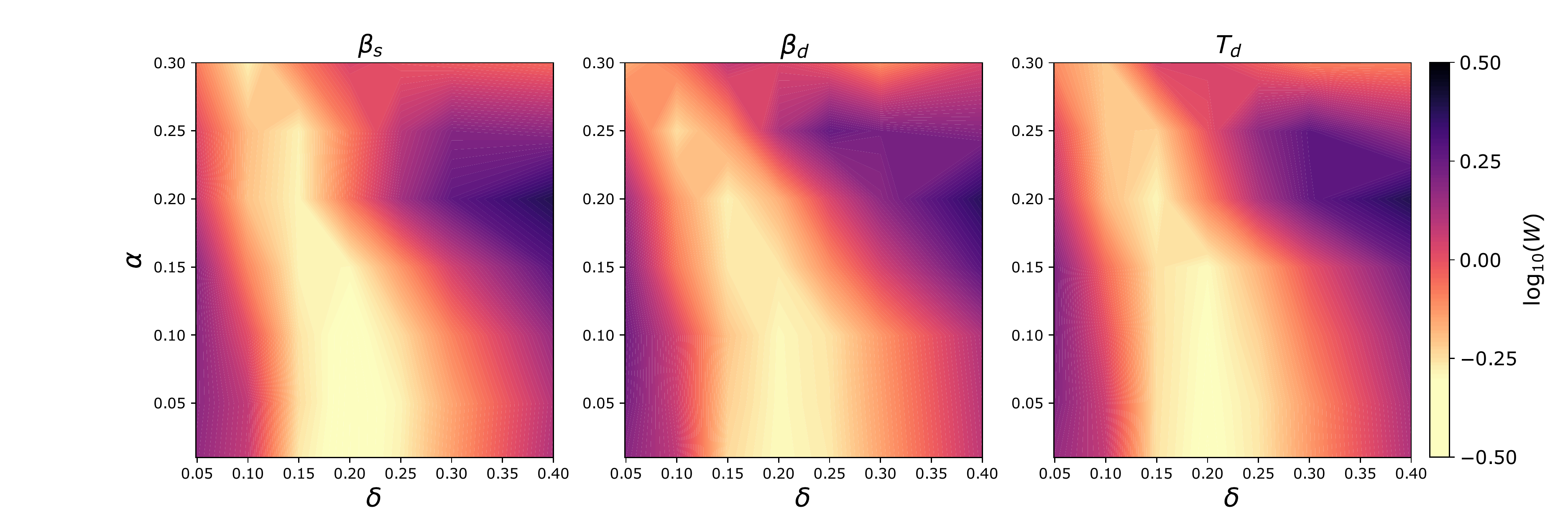}
    \caption{Variance surfaces estimated for  several choices of $\alpha$ and $\delta$, for (left) $\beta_s$, (middle) $\beta_d$, (right) $T_d$. Note that regions of optimality correspond to local minima (lighter regions) in the surfaces.    }
    \label{fig:fg_var}
\end{figure*}

\begin{figure*}
    \centering
    \includegraphics[scale=.47, trim =0cm .1cm 0 4cm,clip=true] {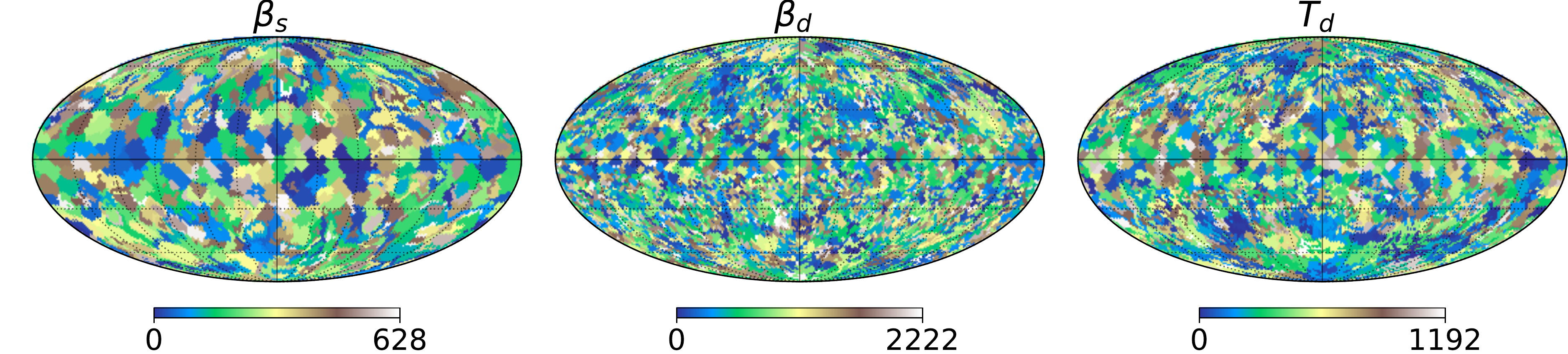}\\
        \includegraphics[scale=.47, trim =0cm .1cm 0 4cm,clip=true] {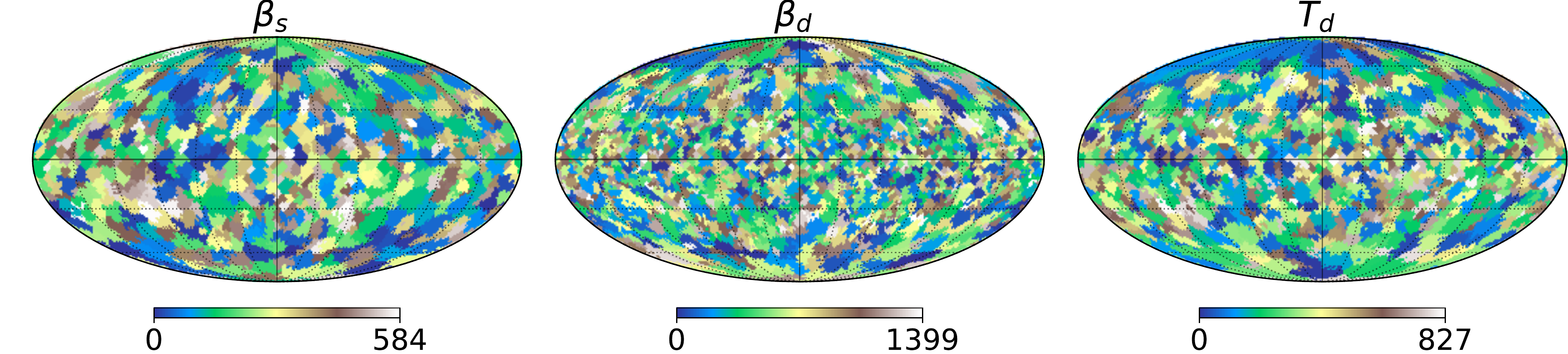}
    \caption{Cluster defined patches obtained from the parameters in the   \texttt{d1s1} model (each color denotes pixels belonging to different clusters). We show Clusters estimated  with $f=0$ (top row)  and with $f=1$ (bottom row).}
    \label{fig:clust_patches}
\end{figure*}
  
\begin{figure*}
    \centering
    \includegraphics[scale=.46, trim =0 0 0 4cm,clip=true ] {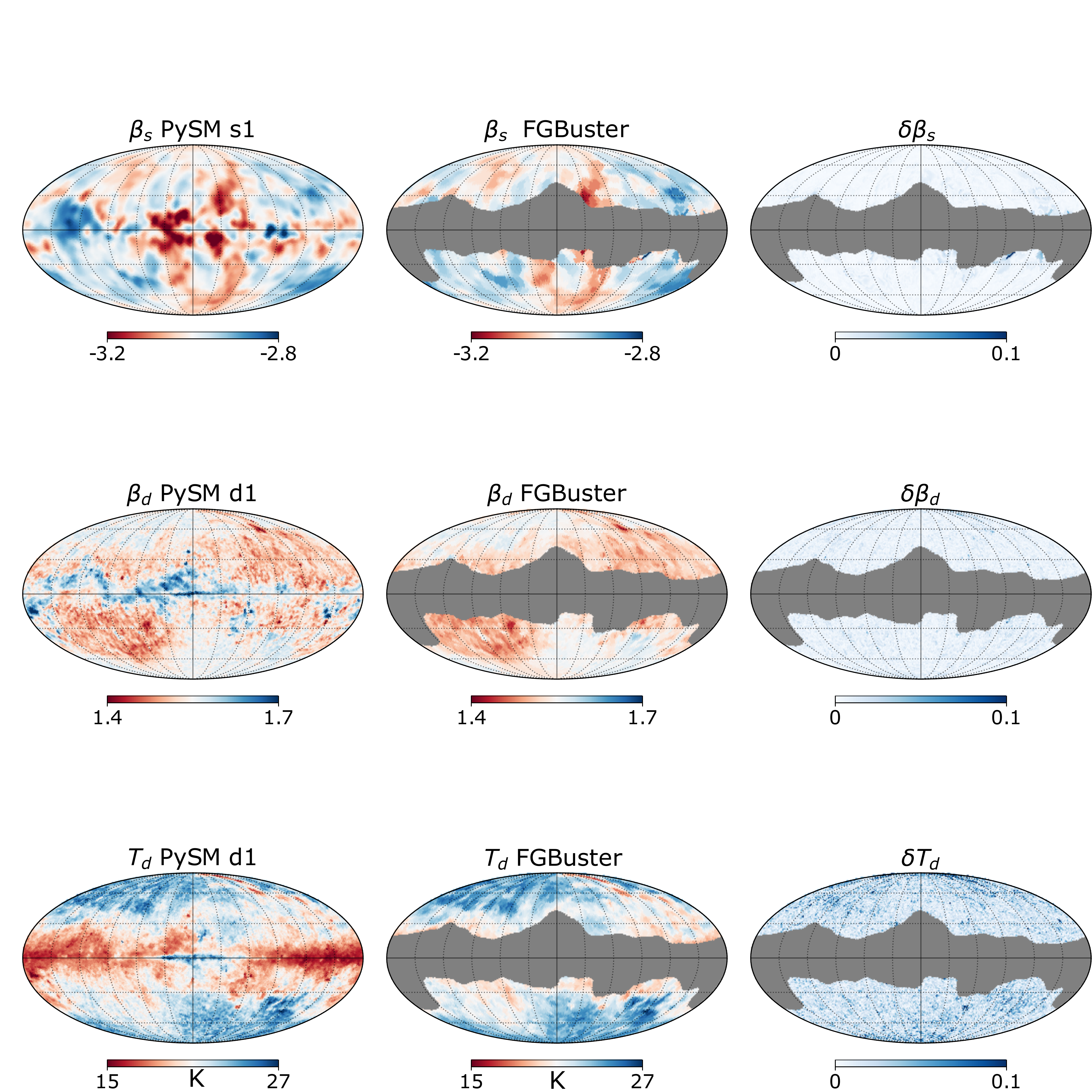}
    \caption{(left column) Foreground parameters from the PySM \texttt{d1s1} model,  (middle column)   parameter maps estimated with FGBuster  on each clustering region (see Fig.~\ref{fig:clust_patches}) using the \lb frequency channels. (right column) Relative residuals on the estimated maps with FGBuster and the input ones from PySM. }
    \label{fig:clust_est}
\end{figure*}
 
\begin{figure}
     \centering
     \includegraphics[width=\columnwidth]{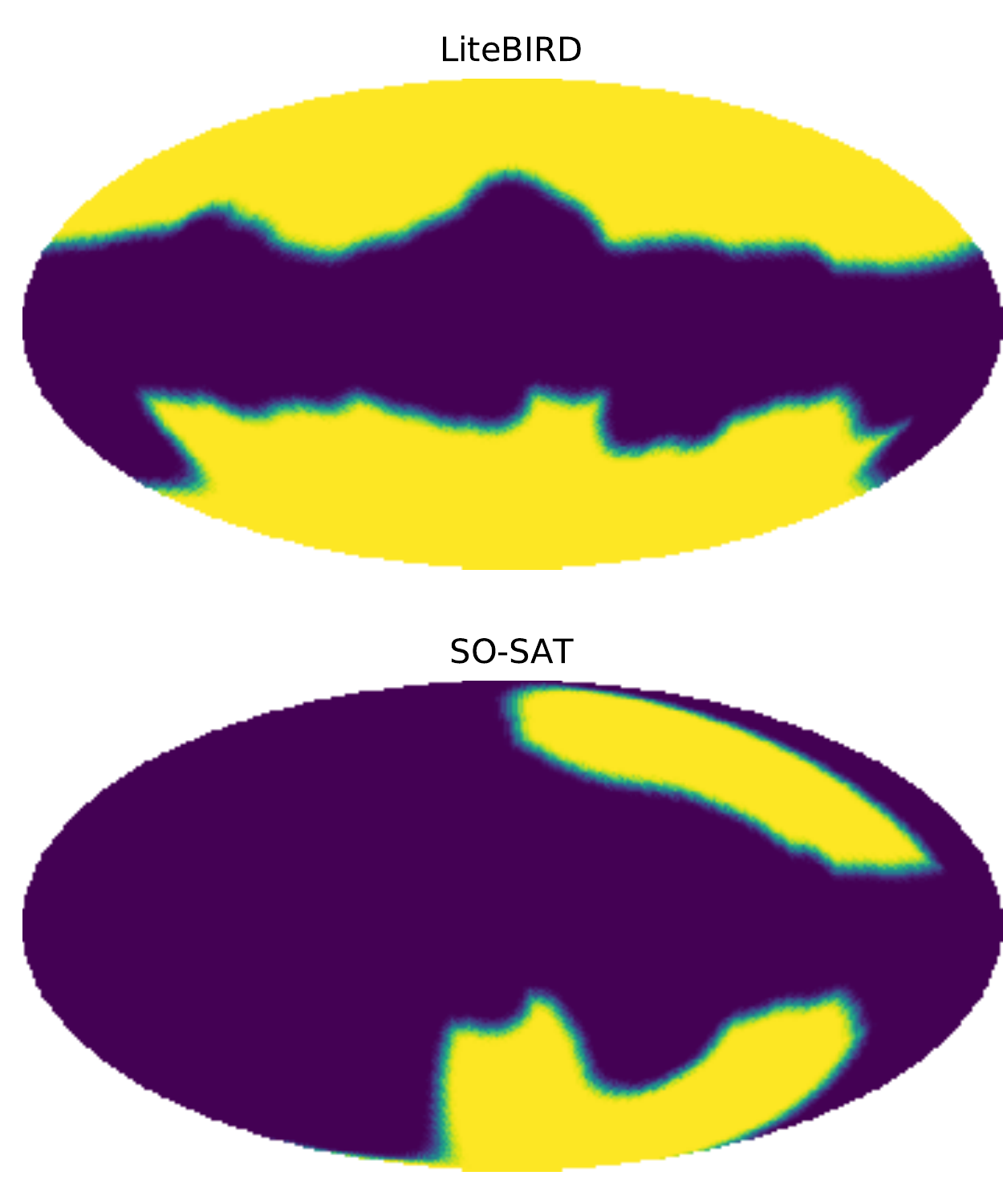}
     \caption{ Masks used for component separation and for the power spectra estimation with \textsc{NaMaster} \citep{Alonso_2019} for \lb (top) and  \sosat (bottom)  encoding respectively    $60\%$ and  $27\%$ of the sky.}
     \label{fig:apomasks}
 \end{figure}

 We  perform parametric component separation   by fitting  for three components: CMB, dust and synchrotron emission. 
 We consider two different combinations of data, to be  representative of the forthcoming experiments aimed at observing primordial $B-$modes:  the Simons Observatory Small Aperture Telescope  \citep[SO-SAT,][]{Ade_2019} from the ground and the \lb space satellite \citep{Sugai_2020}. The component separation is performed  respectively on    $27\%$ and $60\%$ of the sky (see Fig.~\ref{fig:apomasks}) with frequency channels encoding the specifics shown in Table~\ref{tab:specs}. Notice that for \sosat we assume conservatively the \emph{baseline} configuration as described in  \citet[]{Ade_2019}.

We use the Python Sky Model\footnote{\url{https://pysm3.readthedocs.io/}} \citep[PySM]{Thorne_2017} to simulate  full-sky    polarized  emission of thermal dust, CMB and  synchrotron\footnote{The contribution in polarization from AME, CO and  free-free is expected to be very small and it is neglected in the following.}.
 The synchrotron radiation is  commonly parametrized as a power law 
 \begin{displaymath}
I_{\nu,synch} \propto \nu ^{\beta_s}, 
 \end{displaymath}  
 whereas the thermal dust   emission  is  described by  a modified black-body, i.e. 
 \begin{displaymath}
 I_{\nu, dust} \propto \nu^{\beta_d} B_{\nu} (T_d),
 \end{displaymath} 
 with  $ T_d$ being the  black-body dust temperature.  

Among the different models available in PySM, we choose as a reference the \texttt{d1s1} model, which accounts for spatial variation in the synchrotron and dust spectral indices, $\beta_s, \beta_d$ and  in $T_d$. 

The synchrotron spectral index of the \texttt{s1} model is the   \citet[model 5]{Miville_Desch_nes_2008}, estimated by combining the \citet[]{1982A&AS...47....1H} map at 408 MHz  and the map at 23 GHz from WMAP \citep[]{Hinshaw_2009}. The $\beta_s$ map  and its uncertainties $\sigma (\beta_s)$ are available online\footnote{\url{https://lambda.gsfc.nasa.gov/}} at $5$ deg resolution. 

The \texttt{d1} model is derived from the \planck data products\footnote{\url{https://pla.esac.esa.int}} released in \citet{pldiffuse2015}, with  $\beta_d$ and $T_d$ parameters   obtained at about $1$ degree resolution with the COMMANDER component separation algorithm \citet{pldiffuse2015}. 

When multiple frequency channels are simulated with the \texttt{d1s1} model, algorithms relying on parametric component separation would tend to  reconstruct  CMB maps with large    residual foreground bias (commonly referred as \emph{systematic bias}) if a single constant spectral parameter is  fitted for each Galactic component across the whole observed sky, as   the spatial variability of foregrounds is not taken into account in the fit.

On the other hand, the instrumental  noise is responsible for the   \emph {statistical} uncertainties which can be   commonly assessed by running  component separation on  frequency maps encoding   different  Monte-Carlo (MC) realizations of  instrumental noise  and astrophysical signal. 

A partition of the sky where the parameters are fit independently on  multiple regions might mitigate the systematic  bias but increase  the statistical uncertainties   as the fit is then performed on a smaller number  of pixels, hence encoding   a lower signal-to-noise ratio  (SNR) than the case with all the observed pixels~\citep[]{Errard_2019}.

Once the set of parameters is defined together with their associated uncertainties,    we can run the spectral clustering optimization procedure as outlined in Sect.~\ref{sec:spectral}. The   variance  planes for $\beta_s,\beta_d,T_d$  are shown in Fig.~\ref{fig:fg_var}.  
 We show the optimal patches selected in correspondence of  local minimum  of the variance in Fig.~\ref{fig:clust_patches}. 

 We estimate the median value of the number of pixels included in each cluster and we find about $4$ pixels per cluster for $\beta_d$, $8$  for $T_d$  and $16$ for $\beta_s$,   resulting in a typical  angular sizes  for the cluster range from  $\sim 3.5,
\, 7,
\,14$ deg  respectively for $\beta_d$,  $T_d$, and $\beta_s$. The different morphologies and sizes, as well as the different number of clusters, result from a trade-off between  the intrinsic variability due to the  astrophysical emission,  the resolution of each map (see left column of  Fig.\ref{fig:clust_est}) and  the SNR. Specifically, the clustering algorithm finds it harder to agglomerate pixels related  to  very high SNR regions  (e.g. at low Galactic latitudes) with respect to regions with larger uncertainties. In fact, this trend is observed in all the regions: smaller patches  where the SNR is high,  larger patches where it is low.

 We then implement a new functionality in  \textsc{FGBuster}\footnote{\url{https://fgbuster.github.io/fgbuster}}\citep{2016PhRvD..94h3526S}, a package for  parametric component separation, aimed at  fitting  parameters within the set of multiple partitions of the sky. 
 The emission model proposed here is slightly different than in a common parametric setting:  each  spectral parameter is fitted using  data within a given sky region, which can have any shape. For example,  a single $\beta_s$  can be fitted within each region shown in Fig.~\ref{fig:clust_patches} (left), while at the same time  $\beta_d$ is fitted on a different set of regions (see middle panel, Fig.~\ref{fig:clust_patches}). The method is implemented within the \textsc{FGBuster} framework and further details of the implementation will be presented in two companion papers Errard et al. (in prep.) and Poletti et al. (in prep). 
 
 For   a pixel $p$,  the model of emission is given by: 
 \begin{align}
 \small
 m^{p,X}(\nu) &= A^{p,X}_{s} f_{s }(\nu, \beta_s) +A^{p,X}_{d } f_{d }(\nu, \beta_d,T_d) + \nonumber \\
 &A^{p,X}_{cmb} f_{cmb}(\nu) + n ^{p,X}(\nu) , 
 \label{eq:compsep}
 \end{align}
i.e.  a linear combination of the amplitudes of each astrophysical component and the functions $f_{s }(\nu, \beta_s) , f_{d }(\nu, \beta_d,T_d) ,f_{cmb}(\nu) $ encoding the spectral dependence of  synchrotron, dust, and CMB, a  at a given frequency $\nu$ as outlined in Sect.~\ref{subsec:compsep};  $ n(\nu) $ accounts for  instrumental noise coadded to the signal.    Furthermore, we fit for the same spectral parameter both for $X=Q,U$ maps.

 We generate  20  Monte-Carlo realizations of noisy frequency maps with the nominal specifications  of   \lb  and for SO-SAT, see details in Table \ref{tab:specs}. The frequency maps are  produced  with the PySM \texttt{d1s1} model  at \texttt{nside=64} and include both  the input astrophysical polarized signal and the instrumental  noise (assumed to be only white noise).
We show in Fig.~\ref{fig:apomasks} the nominal sky patches where the parametric fit is performed. Notice that we choose a larger $f_{sky} =0.27$ for   \sosat   with respect to the effective $f_{sky} =0.10$ reported  in \citet[]{Ade_2019}. The reason for this choice is mainly motivated by the fact that we want to conservatively assess the performance of this methodology onto larger regions of the sky being thus more susceptible to larger Galactic residuals. This choice further allows us to  compare our results with the ones reported in \citet{2019arXiv190508888T}.

 In Fig.~\ref{fig:clust_est} (middle column), we show the parameter maps estimated in each cluster region with the \lb frequencies  and we evaluate the relative error of this estimate by considering the difference between the PySM parameter maps and the parameters estimated with \textsc{FGBuster} (see right column in Fig.~\ref{fig:clust_est}).
\begin{figure}
    
    \includegraphics[scale=.6, trim=.7cm .8cm .5cm  .5cm , clip=true]{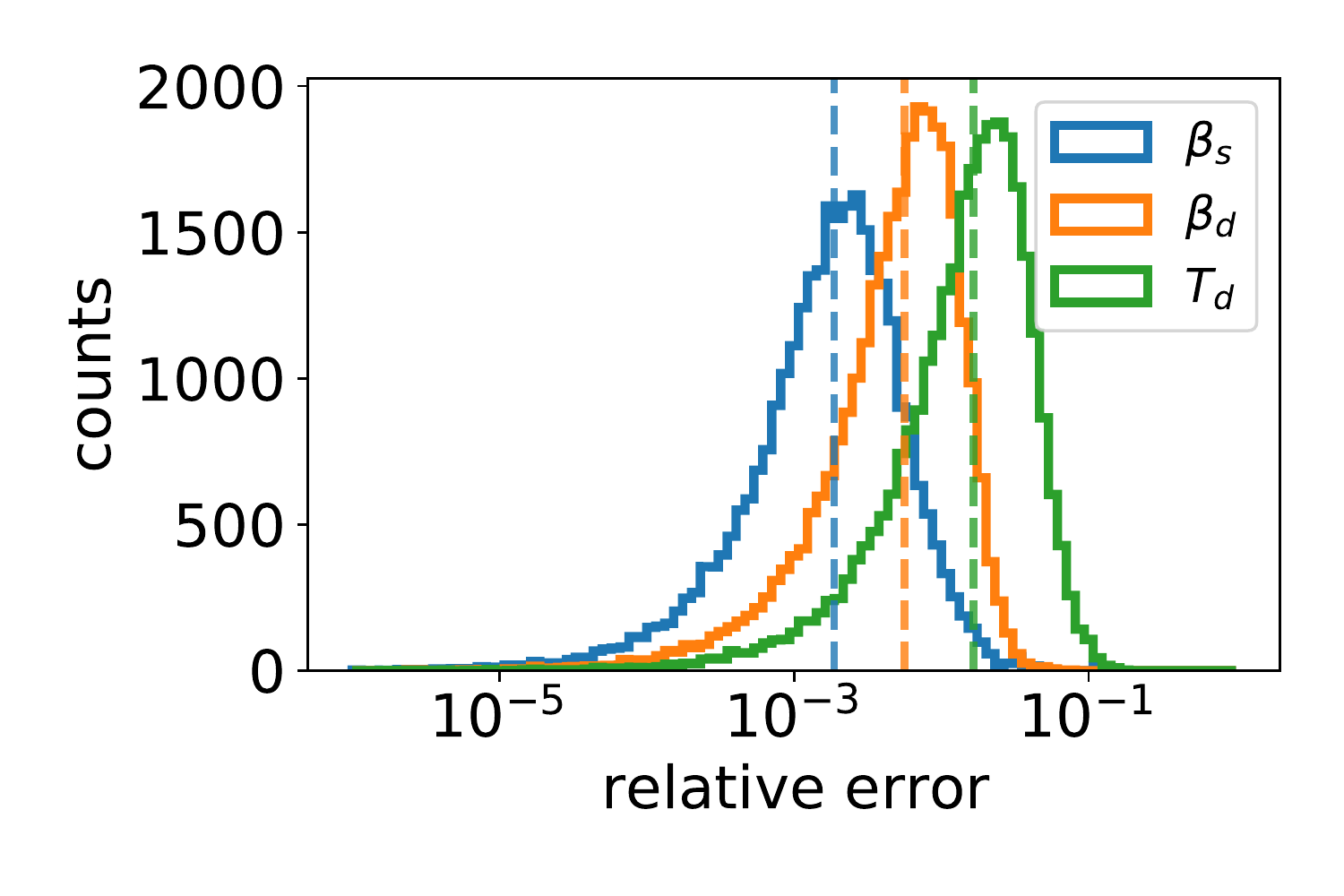}
    \caption{Histograms evaluated from the relative error maps of $\beta_s,\beta_d, T_d$ (right column of Fig.~\ref{fig:clust_est}) respectively in solid blue, orange and green. Shown as vertical dashed lines the median values, namely $0.002, 0.006, 0.017$ respectively for $\beta_s,\beta_d, T_d$.  }
    \label{fig:histo_betas}
\end{figure}

In Fig.~\ref{fig:histo_betas},  we show the distributions of relative errors for the three parameters together with the median values shown as vertical dashed lines to be  $0.002, 0.006, 0.017$ respectively for $\beta_s,\beta_d, T_d$, with the largest errors, $\leq20\%$, found in $T_d$ estimates. This implies that the component separation performed on the cluster-defined patches in Fig.~\ref{fig:clust_patches} (top) recovers  faithfully the spatially varying Galactic components. 
Furthermore, we somewhat expect   the worst estimate to be the one related to $T_d$, since both \lb and \sosat high frequency channels probe frequency regimes  far from the modified black-body peak   ($\nu \sim 600 $GHz), making $T_d$ essentially degenerate with $\beta_d$ at lower frequencies.

\begin{figure*}
    \centering
    \includegraphics[scale=.46 , trim =.80cm 0cm 6cm .0cm,clip=true] {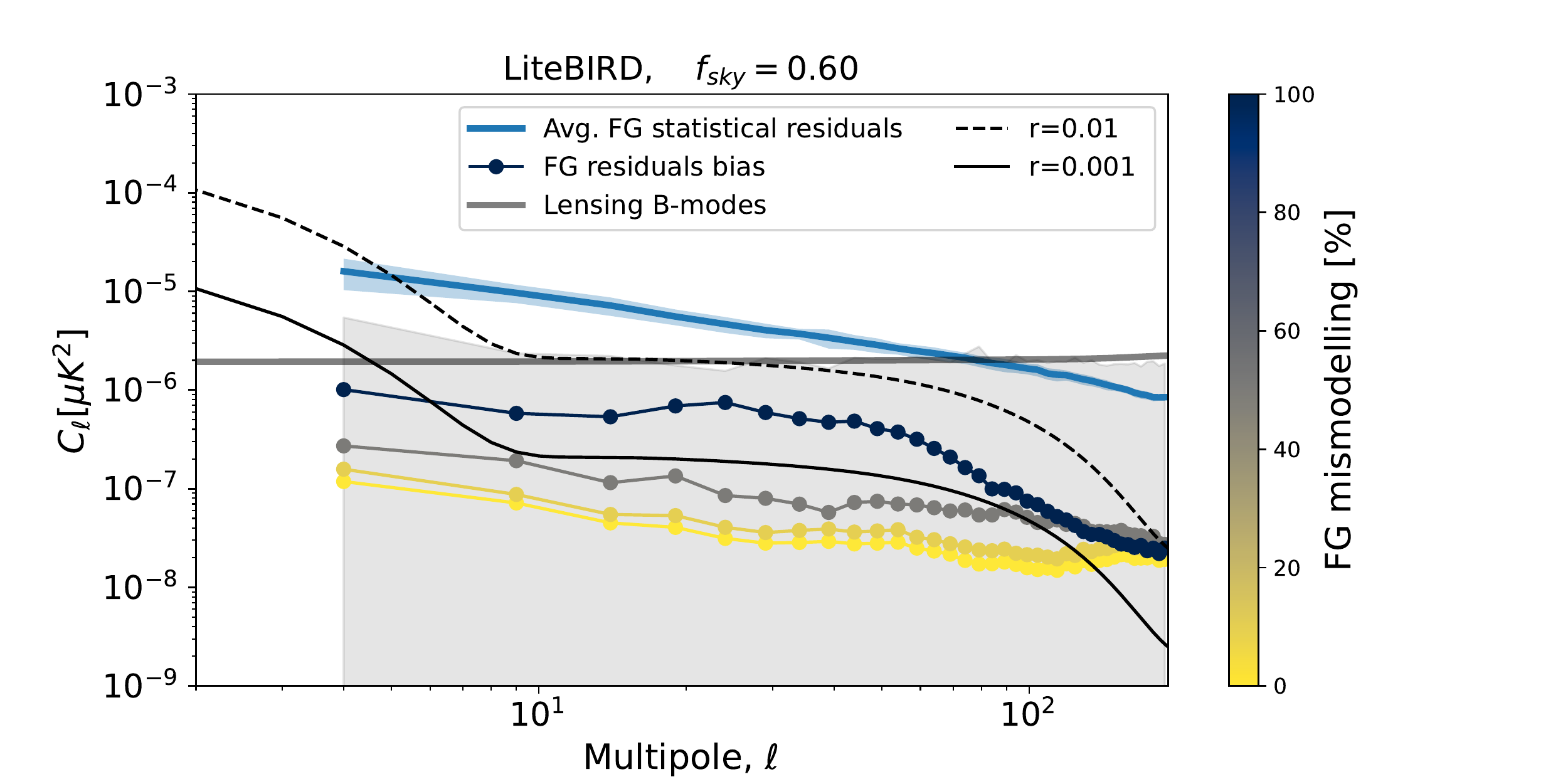}
    \includegraphics[scale=.46, trim =3cm 0cm 3cm .0cm,clip=true] {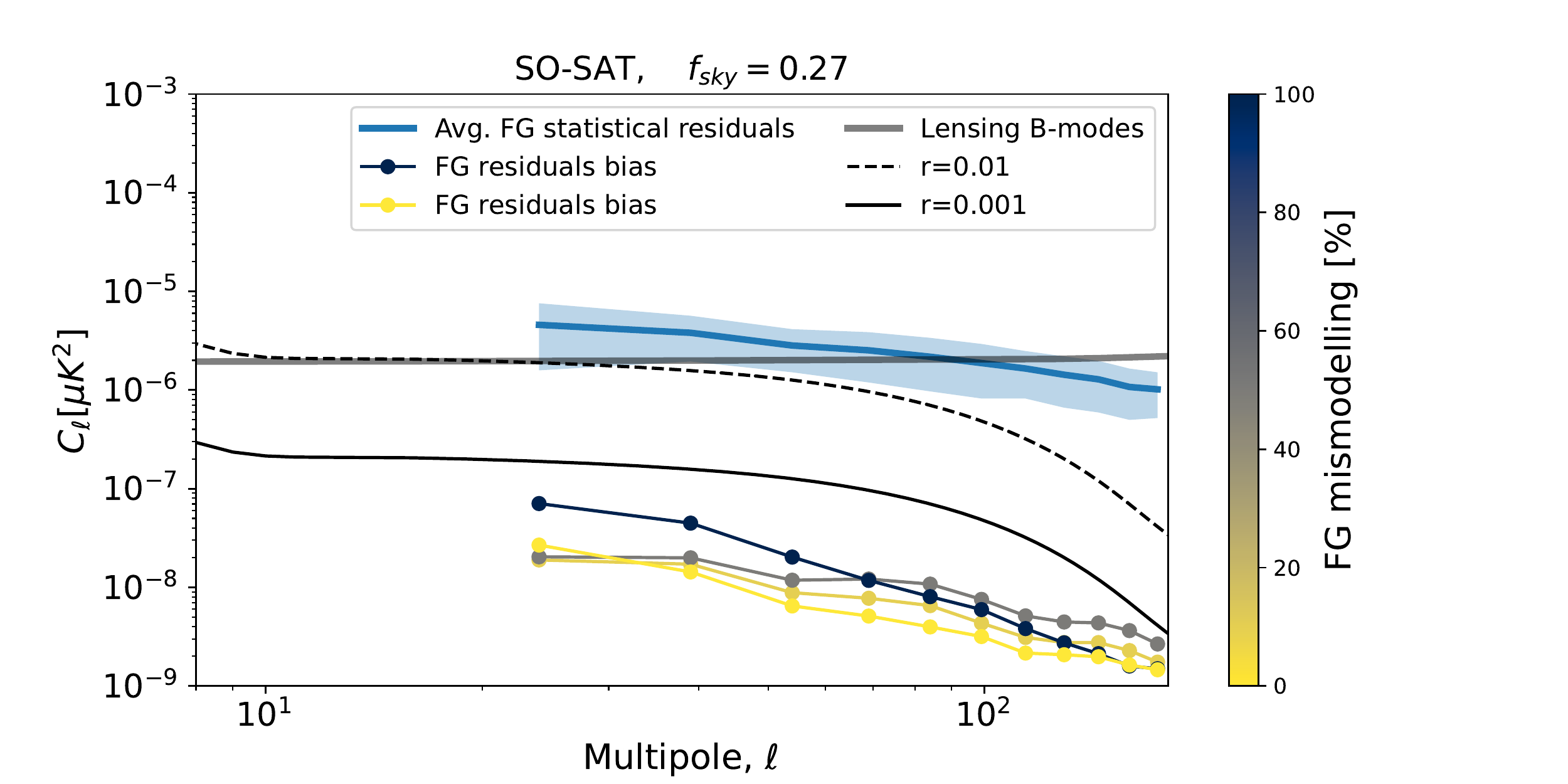}
    \caption{ $B-$mode angular power spectra for the (left) \lb and (right) \sosat cases,  estimated with \textsc{NaMaster} onto the recovered CMB maps from  \textsc{FGBuster} performed onto cluster regions. (solid blue thick) Average of foreground statistical residuals obtained from 20 MC signal+noise   simulations (shaded blue) $1\sigma$ standard deviation of MC simulations, (filled circles)   foreground residual bias estimated with noise-less component separation for different choices of  foreground mis-modelling given by the colorbar.
    The shaded grey area in the left panel indicates two extreme cases: the upper limit is obtained by propagating through FGBuster the $f=1$ mis-modelling to the spectral parameters, the lower limit spectra (not shown here) is nearly equal to zero and it is estimated by propagating in the component separation the $f=0$ case. 
    As a reference we show the primordial B-modes for two different chosen values of tensor-to-scalar ratio, $r=0.01,0.001$ respectively in (dashed black) and (solid black). Lensing B-modes are also shown in (solid thick gray). The spectra in the left and right panels are binned respectively with $\Delta \ell=5$ and  $\Delta \ell=15$. }
    \label{fig:clust_spect}
\end{figure*}
The case shown in Fig.~\ref{fig:clust_est}   is an ideal-limit  case as the patches have been derived by clustring onto the same parameter maps that are used to simulate  the  frequency channels. In order to introduce a \emph{mis-modelling } between the two set of maps, we inject a noise proportional to  the per-pixel  uncertainties on $\beta_s, \beta_d, T_d$,  as : 
 \begin{equation}
 \tilde{X} = X + f \,\,N_{w}[\sigma(X)] , 
 \label{eq:mis-model}
 \end{equation}
 where $X=\beta_s, \beta_d, T_d$; $N_{w}[\sigma(X)]$ is  a random Gaussian noise map with zero mean and width given by  the uncertainty map, and $f$ is a weighting  constant factor,   ranging from 0 to 1. 
 
 The   $f=1$ case  represents a pessimistic case where the foregrounds are fully mis-modelled given the uncertainty budget in the spectral parameters and $f=0$ represents the ideal case   without any  mis-modelling.  

 For each mis-modelling case, we estimate the statistical residuals by performing component separation on maps encoding  20 MC independent realizations of instrumental noise and astrophysical signal.  The systematic bias is instead estimated by running component separation on noiseless maps.
 
We therefore run the spectral clustering algorithm on the mis-modelled parameter maps $\tilde{\beta}_s, \tilde{\beta}_d, \tilde{T}_d$ for several values of $f$.{ E.g. in Fig.~\ref{fig:clust_patches} (bottom), we show the cluster obtained from parameter maps with   maximum mis-modelling ($f=1$). We note a trend to have smaller number of  clusters with similar sizes and more uniformly distributed with respect to the $f=0$ case for similar values of $\alpha, \delta$. This is particularly noticeable for the dust parameters since they are provided with a higher resolution than the synchrotron one. We conclude that injecting the mis-modelling noise into the spectral parameter maps tends to homogenize the typical size of the patches.   } 

We then  perform the component separation separately for  \lb and \sosat channels (simulated without mis-modelling)   onto the cluster patches derived with the mis-modelling. This procedure allows us to further  set requirements on the   mis-modelling  given the residual bias we observe in the recovered CMB map for different values of $f=[0,0.1, 0.5,1] $.
 
 As the output maps of \textsc{FGBuster} are estimated on a partial sky (see Fig.~\ref{fig:apomasks}),  we estimate the power spectra in the observed regions with \textsc{NaMaster} \citep{Alonso_2019} to correct the power spectra for the $E-B$ leakage introduced by masking. We show in Fig.\ref{fig:clust_spect} the angular power spectra of CMB $B-$mode  polarization anisotropies as estimated from the  recovered  map outputs of \textsc{FGBuster}. 
 
 We notice that the statistical residuals are not affected by the mis-modelling since the instrumental noise is the dominant contribution to the post-component separation residuals. On the other hand, the mis-modelling is clearly visible when we assess the systematics bias in the residual maps. In particular,  we observe an increase of  the systematic bias proportional to $f$, indicating   how  the partition is less and less  representative with increasing values of $f$ of the underlying Galactic foreground emission. However, even with the largest mis-modelling scenario $f=1$,  the residuals increase by only an order of magnitude.  As expected,  larger $f_{sky}$ (including lower Galactic latitudes)  yields larger  residuals in the power spectra, this is the reason why we observe  higher residuals  for  \lb  with respect to the  \sosat case).

  {Furthermore, to show the effective gain of performing parametric component separation on patches defined with clustering, we consider two limiting cases where we estimate the systematic residuals by evaluating  eq.~\eqref{eq:compsep} for \lb  with the maps of $\beta_d, \beta_s, T_d$ for $f=0, 1$ .  Obviously, the case $f=0$ gives numerically zero residuals as we  provide the exact solution to the parametric fit (i.e. the same spectral parameter maps with which we have constructed the frequency channels). On the other hand, the case $f=1$ resembles the extreme case where the mis-modelling noise is fully propagated through  the component separation pipeline. In Fig.~\ref{fig:clust_spect} (left),  we show as a shaded gray area the two extreme cases, with $f=1$ being the largest error that can be introduced with the mis-modelling defined in eq.~\eqref{eq:mis-model} (the $f=0$ is not shown for graphical purposes). The fact that the power spectra obtained with clustering are well within this range, quantifies the overall benefit of performing the parametric fitting with the clusters even if clusters are obtained with the largest  mis-modelling case ($f=1$).   }
 
 {Finally, we notice in Fig.~\ref{fig:clust_spect} that all the $B-$mode  systematic residual spectra tend to converge  to a certain {residual  level} at  around $\ell \gtrsim 60$. Hereafter, we refer to it  as a \emph{partition noise} being  essentially related to the combination of the  typical angular scales of the cluster patches. This results as a ``noise'' residual term in the power spectrum, as that prevents to estimate the  variability of the  spectral parameters   at multipoles larger than this scale. }
 

\subsection{Clustering applied on HI clouds maps.}\label{sec:nc}

 In the previous section, the component separation is performed on partitions derived from  the same maps used in the parametric modeling, i.e. $\beta_s, \beta_d, T_d$.  However, this relies on the  assumption that spectral parameter maps, from which we derive the cluster patches,  are not contaminated   by  other Galactic  (or extra-galactic) foregrounds. Indeed, this is commonly  the case for both the  thermal dust emission,  mainly affected by the  Cosmic Infrared Background (CIB)  residuals   and  for synchrotron contaminated  by  free-free and AME. 
 
 Moreover, given the fact that  the use of ancillary datasets for foreground modeling has been proposed by a number of works (e.g. \citet[]{hi4pi2016,Clark2018}),  we wish to explore whether such an ancillary dataset can in principle yield accurate enough results for CMB parameter estimation purposes. To this end, we use a complementary dataset that traces foreground dust: HI emission. In particular, we make use of the parameterization of line-of-sight complexity in the dust distribution presented by \citet[]{Panopoulou2020}. We briefly explain the dataset and our post-processing of it in the following.

\citet{Panopoulou2020} measured the number of clouds along the line of sight that appear as distinct peaks in HI spectra.
They present a peak-finding algorithm that makes use of the HI4PI survey \citep[]{hi4pi2016} over the high Galactic latitude sky and outputs a measure of the number of clouds in each pixel. They introduced a robust measure of the number of clouds expected to contribute to the dust emission, which takes into account the relative contribution of the dust from each cloud along the line of sight, defined as $\mathcal{N}_c$:

\begin{displaymath}
\mathcal{N}_c = \sum _{i=1} ^{N_{\mathrm{clouds} }} \frac{N_{\mathrm{HI}}^i }{N_{\mathrm{HI}}^{\mathrm{max}} } 
\end{displaymath}
 where $N_{\mathrm{HI}}^i$ refers to the column density of the $i$-th cloud along the sightline and  $N_{\mathrm{HI}}^{\mathrm{max}}$ is the highest value of  column density for a cloud in the same direction. We therefore expect non-integer values for $\mathcal{N}_c$ especially  when there is an imbalance in the column density  for clouds along a sightline.   
 The $\mathcal{N}_c$  map was estimated by segmenting the sky into large \emph{superpixels}  at \texttt{nside=128} (corresponding to $\sim 30 $ arcmin), with each superpixel encoding  64 pixels of the HI4PI map.
Recently, \citet{Pelgrims2021} adopted the maps publicly released by \citet{Panopoulou2020} to estimate the line-of-sight frequency decorrelation of dust polarization in the \planck maps. For the purposes of our clustering analysis, we require maps at \texttt{nside=32} and \texttt{nside=64}. We thus repeat the analysis in \citet{Panopoulou2020} to produce maps at these lower \texttt{nside}s, instead of downgrading the publicly available  maps, as recommended by the authors.
The publicly available $\mathcal{N}_c$ maps did not include uncertainties. As our clustering relies on the use of uncertainties, we calculate the per-pixel uncertainty in the $\mathcal{N}_c$ maps that we use as described in Appendix~\ref{appendix:Nc}.

As discussed in \citet[Sect. 5.3.1] {Panopoulou2020},  the derived maps do not present  any imposed spatial coherence at scales above the superpixel size, leading  to the eventuality of  discontinuities  between neighbouring  pixels. 
However, clustering methodologies presented in this work, can implement the spatial coherence above a given pixel scale. 

\begin{figure*}
    \centering
    \includegraphics[width=1.5\columnwidth, trim=0 0 0 0cm, clip=true]{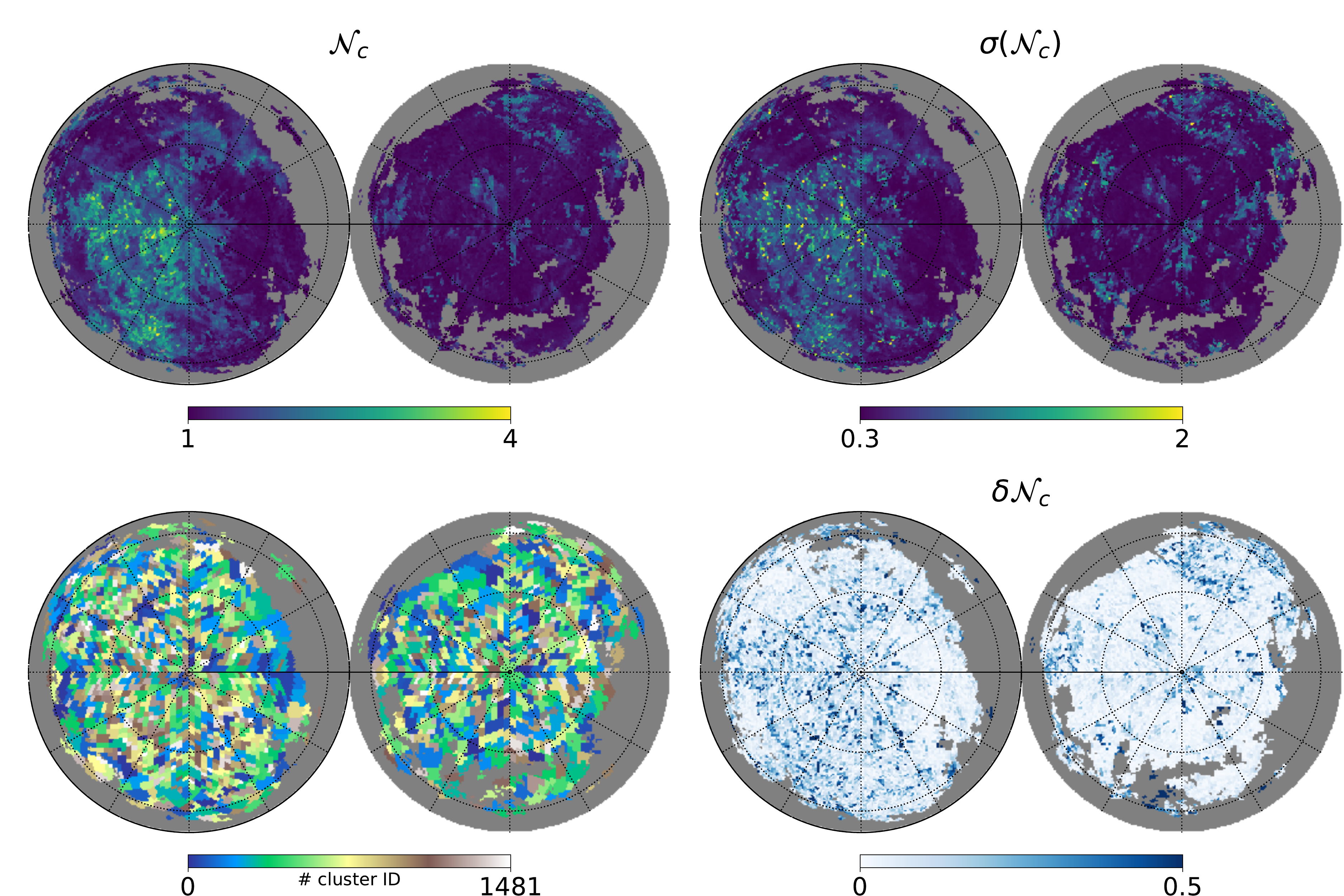}
    \caption{(top) Maps of $\mathcal{N}_c$ (left) and its uncertainties $\sigma(\mathcal{N}_c) $ (right) evaluated as in  eq.~\eqref{eq:nc_err}. (bottom left) Partition of $\mathcal{N}_c$; each color identifies a different clustering region. (bottom right) Relative errors $\delta \mathcal{N}_c$  estimated with  (eq.~\eqref{eq:rel_res}).}
    \label{fig:nc_clus}
\end{figure*}

We employ the $\mathcal{N}_c$ and the associated uncertainties as in~\eqref{eq:nc_err} with maps at \texttt{nside=32} (shown in top panel of Fig.~\ref{fig:nc_clus}) and run the spectral clustering methodology outlined in Sect.~\ref{sec:spectral}. We then evaluate the total variance of the partition in a similar grid to the one shown in Fig.\ref{fig:fg_var} and find the minimum variance to be at  $\alpha =0.10$ and $\delta=0.20$, corresponding to  $K=1481$  clusters.

\begin{figure*}
    \centering
    \includegraphics[scale=.5, trim=0 0cm 0 0cm, clip=true]{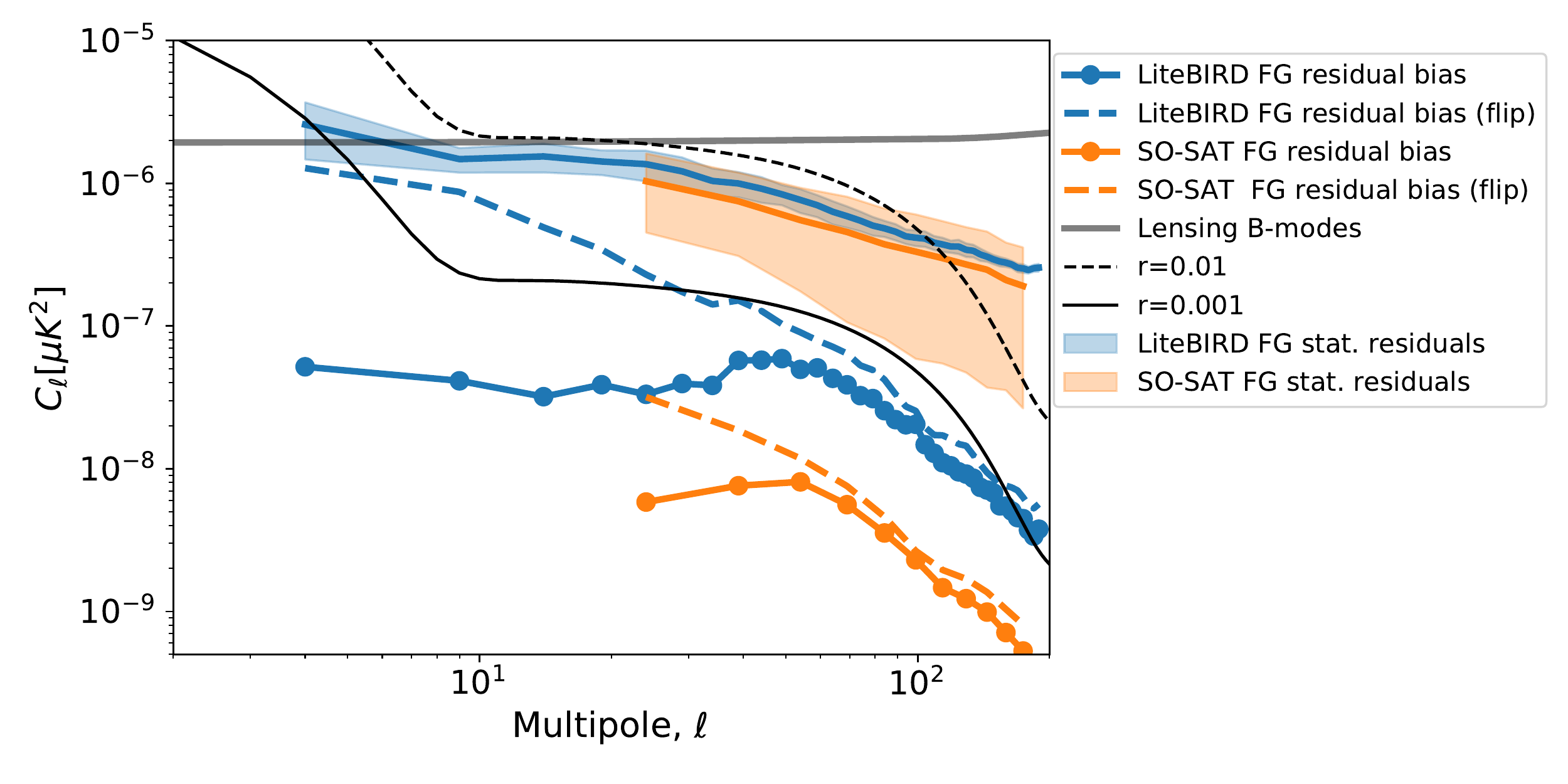}
    \caption{$B-$mode residuals   from the recovered CMB maps output of \textsc{FGBuster} performed within multiple regions defined by clustering of $\mathcal{N}_c$  map for (blue) \lb  and (orange) \sosat frequency channels, respectively on $f_{sky} =0.35$ and $0.22$. (filled circles) Foreground residual bias performed on noiseless simulations, (shaded area) $1\sigma$ standard deviation of statistical residuals evaluated from 20 MC signal and noise simulations  (the average of the 20 MC realization is shown as solid line). (dashed) Foreground residual bias estimated from the component separation run  performed by flipping the map of $\mathcal{N}_c$ clusters North-to-South. These spectra show an excess at lower multipoles, demonstrating that the $\mathcal{N}_c$ map provides useful constraints for parameter estimation.   }
    \label{fig:nc_spectra}
\end{figure*}

To evaluate how well the clustering morphologies track the input  map, we estimate the median value of pixels of $\mathcal{N}_c$ belonging to the same clusters and produce a \emph{binned}  map, $\mathcal{N}^{bin}_c$. We can then estimate the relative error  defined as 
\begin{equation}
\delta \mathcal{N}_c= |\mathcal{N}_c-  \mathcal{N}^{bin}_c|/\mathcal{N}_c.
\label{eq:rel_res}
\end{equation}
The bottom row of  Fig.\ref{fig:nc_clus}, shows in the left panel the partition  estimated with spectral clustering    and in the right one the relative errors.
We observe that especially around the North and South Galactic Poles ($90^\circ<|b|<60^\circ$)  the clustering optimization tends to  produce smaller clusters, essentially encoding few pixels per cluster. On the contrary,   at intermediate Galactic latitudes, clusters contain a larger  number of pixels.  This might be related to the fact that the $\mathcal{N}_c$ maps are produced on too large pixelization  (\texttt{nside=32})  resulting in an overall reduction of the uncertainty budget and a consecutive increase in SNR of the estimates. As already stated in Subsection~\ref{subsec:compsep},  in the presence of very high SNR features, the optimal partition leads to \emph{pixel-sized } clusters.  We will  devote a further investigation on this respect in a future work.

Notice that in the regions where the uncertainties are small, the error is below $\sim 10\%$.  Vice versa, in large uncertainties regions the relative error can be as much as $50\%$ although those regions are  localized  mainly in the North Galactic Hemisphere. However, we remark here that these larger errors should not be compared with the ones shown in Fig.~\ref{fig:clust_est} as larger  errors on $\mathcal{N}_c$ do not necessarily translate on larger residuals on the recovered dust parameters.    

As already mentioned in \citet[]{Panopoulou2020}, one of the interesting applications of the  map of number of clouds per line of sight   is to inform realistic models of polarized dust emission for parametric component separation methods as the one we adopt in Sect.~\ref{subsec:compsep}.  In this specific case, we utilize  the patches derived from the $\mathcal{N}_c$  map (bottom left panel in Fig.~\ref{fig:nc_clus})   to perform the component separation estimation of dust  and CMB (simulated at \texttt{nside=64}). In fact, since  the  $\mathcal{N}_c$  map traces the dust emission, we do not include the synchrotron as well as all the $\nu<90 $ GHz   frequency channels in this analysis.  This helps in  better singling out the effects of residuals  due to the dust component  only. 

Furthermore, to run \textsc{NaMaster} we regularize  the HI4PI  observation area ($f_{sky} =0.55\%$)  by apodizing it, yielding to  $\sim 0.35 \% $ of the sky . For the run  with \sosat we combine this mask with the \sosat one shown in Fig.~\ref{fig:apomasks} (bottom). The combined observation patches results in a  further reduction,   $f_{sky} \sim 22\%$.

In Fig.~\ref{fig:nc_spectra}, we show the $B-$mode power spectra of the residuals in the recovered CMB map estimated with \textsc{NaMaster}. We observe that the statistical residuals are reduced by about one order of magnitude compared to the ones shown in Fig.~\ref{fig:clust_spect}. This is mostly due to the fact that we have excluded the low frequency channels that have lower sensitivity (see Table~\ref{tab:specs}) yielding  larger  post-component separation noise  when taken into account.
  Interestingly, the residual bias achieves levels comparable to $ r \lesssim 10^{-3}$.
  It therefore seems that the parameters from the component separation performed on the dust-emission-agnostic $\mathcal{N}_c$-based sky partitioning are accurately representing the underlying signal. 
  
  One might wonder whether this dust-emission-agnostic clustering constitutes a real benefit  in parameter estimation or it is only related to the fact that we are simply partitioning the northern and southern polar caps with small patches.  In other words: \emph{is there intrinsic value in the information contained in the number-of-clouds map?}
  
  To answer  this question, we disentangle the sky partition from the underlying physical information by flipping the cluster coordinates North to South.
  However, since the North and South Polar caps have different footprints, we firstly flip the map and then apply the regularized mask to  the $\mathcal{N}_c$  map. Encoding the mask on a smaller fraction of sky ensures that  each pixel  in the flipped map falls within the $\mathcal{N}_c$ map footprint. 
 
  We then perform the parametric fit with the same combination of frequency channels as before but  with the flipped cluster regions. The thick dashed lines shown in Fig.~\ref{fig:nc_spectra} indicate  the power spectra of systematic residuals from the output of this component separation run and we notice a remarkable increase of the residuals especially at lower multipoles both for \sosat and \lb.

A visual inspection of the relative error and standard deviation maps  estimated in both cases for $T_d$ and $\beta_d$, reveals  that both the  error and the standard deviation  (particularly in $T_d$) increase  when we consider the case with flipped  regions.  This also indicates that the bias we observe in Fig.~\ref{fig:nc_spectra}  is mostly due to  wrong estimation of the $T_d$ parameter. 
 
 This is not unexpected since $\mathcal{N}_c$ has been shown to be correlated with $T_d$ \citep{Panopoulou2020}. The higher values of $\mathcal{N}_c$  coincide with regions where the emission from two physically distinct families of clouds overlap (Low Velocity and Intermediate Velocity clouds). One family of clouds resides at larger distances from the Galactic plane and is found to have higher dust temperatures, presumably due to dust shattering as it falls onto the mid-plane \citep{Planck2011}. Therefore, a mis-modelling of the cloud population  (e.g. by flipping the north-south  map) naturally leads to a wrong estimate of $T_d$.  
 We thus can conclude that partitioning the sky observations with patches inferred from  the $\mathcal{N}_c$  map can improve the performances of parametric fitting methodologies. 
 
 Finally, we observe also for this case  the effects of   a {partition noise} in the power spectra: with  $B-$mode residuals    in Fig.\ref{fig:nc_spectra}  (dashed lines and filled circles) converging at around $\ell \sim 60$. 
 A  partition with  smaller patches, as the one shown in Fig.~\ref{fig:clust_patches}, results into shifting to higher multipoles  the contribution due to this partition noise.
 \citet{Grumitt_2020} proposed a hierarchical   approach  to overcome exactly this limitation and we plan to integrate  it in a  future work. 
 
 \subsection{Estimates on $r$}

 We recall the reader that to specifically assess  the bias introduced by foregrounds, we do not include the CMB emission in the  simulated frequency maps.  However, we instead fit for  CMB   in the parametric component separation. 
 Given the $B-$mode power spectra shown  in Fig.~\ref{fig:clust_spect} and \ref{fig:nc_spectra}, we estimate the value of residual bias in terms of $r$ by evaluating the   likelihood   $\mathcal{L}(r)$  in the  binned multipole domain, $\ell_b$, given the fact that the 
 power spectra have been corrected by the mode coupling with   \textsc{NaMaster}. 
 
The likelihood function reads as:
\begin{equation}
\ln \mathcal{L}(r) = \sum_{
\ell_b=\ell_{\rm min} }^{\ell_{\rm max}} \ln \left(  -f_{  sky} \Delta \ell \frac{2\ell_b+1 }{2}\left[\frac{\hat{C}_{\ell_b} } {C_{\ell_b}} +\ln C_{\ell_b}\right] \right),
\label{eq:global-likelihood}
\end{equation}
where $\ell_{\rm max}=200$ ,  $\ell_{\rm min} =2 \, ( 30)$ and  $\Delta \ell=5 (15)$  for \lb (\sosat) and   $\hat{C}_{\ell_b}$ ($C_{\ell_b}$) is the measured (modeled) B-mode power spectrum. As we are interested in estimating the bias of systematic residuals we assume   the measured $B-$mode spectrum to encode  all the contributions but the primordial one ($r=0$), i.e. :
\begin{equation}
\hat{C}_{\ell_b} =C_{\ell_b}^{\rm lens} + \langle C_{\ell_b}^{\rm fg, tot}\rangle ,
\label{eq:cl-measured-in-likelihood}
\end{equation}
where  $\langle C_{\ell_b}^{\rm fg, tot}\rangle $ is the  expected statistical noise power spectrum residuals after component separation obtained by averaging the    spectra  of 20 MC signal+noise simulations \citep[][]{Errard_2019}  and  $C_{\ell_b}^{\rm lens}$ is the lensing B-mode power spectrum . 
The modeled power spectrum is given as:
\begin{equation}
C_{\ell_b} = r C_{\ell_b}^{\rm tens} +C_{\ell_b}^{\rm lens}+ \langle C_{\ell_b}^{\rm fg, tot}\rangle -  C_{\ell_b}^{\rm fg, sys},
\label{eq:cl-model-in-likelihood}
\end{equation}
where $C_{\ell_b}^{\rm tens}$ is the tensor mode with $r=1$ and  $C_{\ell_b}^{\rm fg, sys}$ is the spectrum encoding the systematics bias due to mismodelling the foreground emission. 

Thus, the bias on $r$, $\Delta r$ is defined as the value that maximizes the likelihood function:
\begin{equation}
\left. \frac{d \mathcal{L}(r)}{dr}\right|_{r=\Delta r}=0, 
\label{eq:r-bias-definition}
\end{equation}
and the error on   $r$, $\delta r$  is defined as the value covering the 68\%
area of the total likelihood function, i.e.:
\begin{equation}
\frac{\int_0^{\delta r} \mathcal{L}(r)dr}{\int_0^{\infty}  \mathcal{L}(r)dr} = 0.68.
\label{eq:r-total-error-definition}
\end{equation}

The values reported in Table~\ref{tab:r_estimates} are extensively discussed in the following section.

\section{Discussion}\label{sec:discussion}

We devote this section to discuss in more detail the results presented in Section~\ref{sec:results}. We estimate how the bias from  systematic residuals  for  different choices of the mis-modelling parameter $f$ affects the detection of   primordial CMB $B-$modes and compare with the requirements for \sosat and \lb.

\begin{figure}
     \centering
     \includegraphics[trim=.3cm 0 .3cm 0,clip=true, width=0.95\columnwidth]{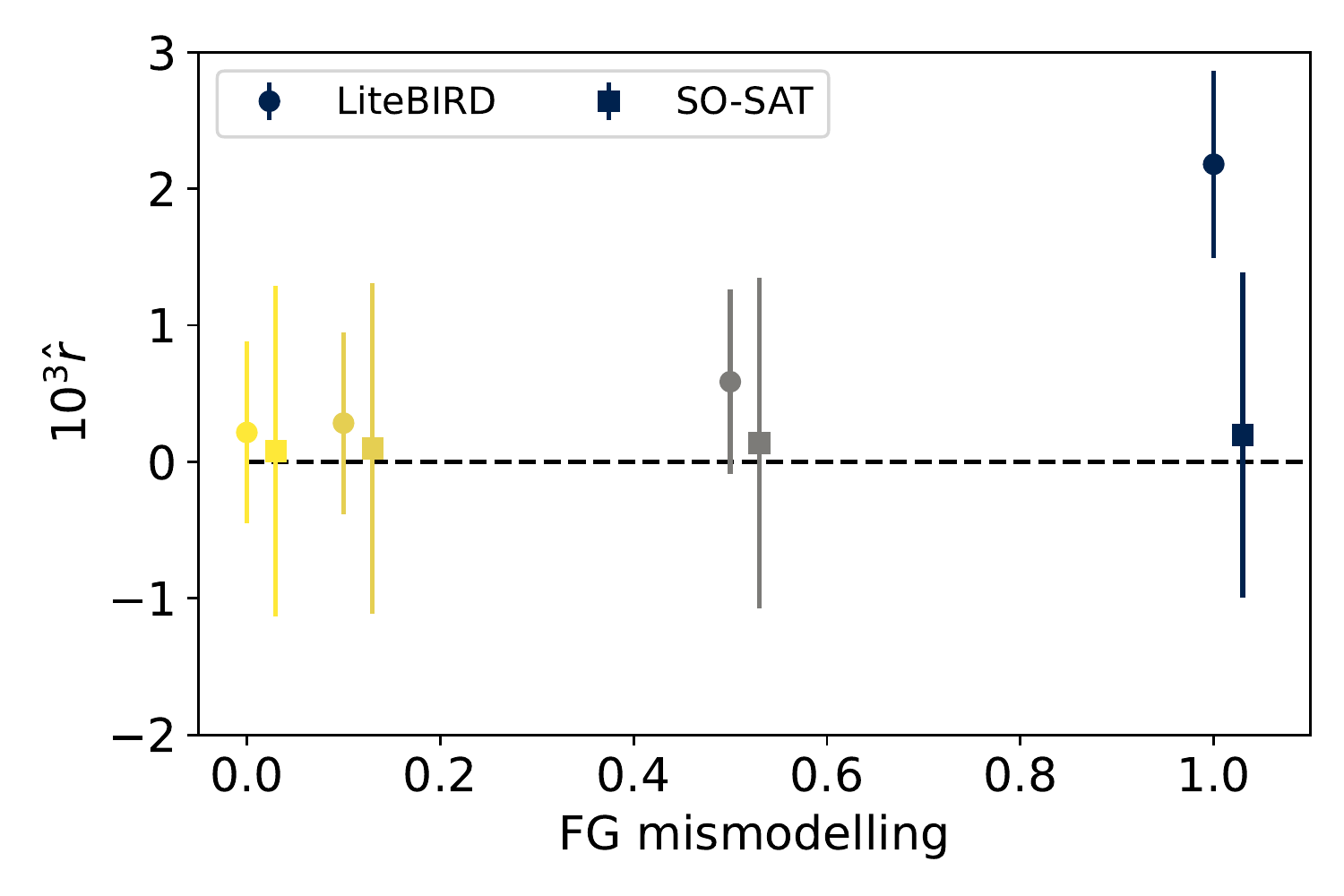}
     \caption{$r$ estimates  and the accompanying uncertainties  evaluating the likelihood in eqs.~\eqref{eq:global-likelihood}, \eqref{eq:r-bias-definition} and \eqref{eq:r-total-error-definition}. We adopt  the same color scheme as in Fig.~\ref{fig:clust_spect} to indicate different amounts of mis-modelling.}
     \label{fig:r_estimates}
 \end{figure}

\noindent As reported in Table~\ref{tab:r_estimates} and shown in Fig.~\ref{fig:r_estimates},  the effect of foreground mis-modelling is visible in terms of  systematic  bias on   $r$ estimate but slightly affects the uncertainties (which are instead driven by the sensitivity of the experiment, $f_{sky}$, frequency coverage, etc\dots).  
 Although the  bias for \lb is always   $<10^{-3}$ for most cases, we observe an increase proportional to the mis-modelling. Without further optimizing the Galactic mask for polarization the values in the top row of Table~\ref{tab:r_estimates}  show that we can meet the latest  \lb  requirements \citep[$\delta r <10^{-3} $, ][]{Sugai_2020}  with a mis-modelling $f\leq 0.5$.
 
 \noindent We notice that the   bias on  $r$ obtained in this work  for \sosat is  slightly lower than the  reported  values in \citet[][Table 4]{Ade_2019}, indicating  a net improvement thanks to this technique. {The larger uncertainties and the lower systematic bias observed in \sosat,  with respect to \lb ones, are mostly due to the smaller  coverage both in frequency and in the angular scales observed.} Moreover, the $r$ estimates for  \sosat  are assessed by evaluating the likelihood in a limited range of $\ell$ ($30<\ell<200$). Vice versa, \lb benefits of the full-sky angular range  allowing  also the large angular scales to be included in the likelihood ($2<\ell<200$).

 \begin{table}
     \centering
     \begin{tabular}{lcc}
     \hline 
     Clustering on  $\beta_s, \beta_d, T_d $ &&\\ 
         FG Mismodelling $f$ &   \lb & \sosat \\
    \hline
$0$& $ (0.22\pm 0.66) \times 10^{-3}$ & $(0.08\pm 1.21)\times 10^{-3}$ \\  
$0.1$& $ (0.28\pm 0.66) \times 10^{-3}$ & $(0.10\pm 1.21)\times 10^{-3}$ \\  
$0.5$& $ (0.59\pm 0.67) \times 10^{-3}$ & $(0.14\pm 1.21)\times 10^{-3}$ \\  
$1.0$& $ (2.18\pm 0.68) \times 10^{-3}$ & $(0.20\pm 1.19)\times 10^{-3}$ \\   
\hline
Clustering on  $\mathcal{N}_c$   & $(0.09\pm 0.26) \times 10^{-3}$ &  $(0.04\pm 0.16) \times 10^{-3}$ \\ 
\hline 
     \end{tabular}
     \caption{Estimated values for the bias  and the uncertainty on $r$   obtained by evaluating the likelihood in eqs.~\eqref{eq:global-likelihood}, \eqref{eq:r-bias-definition} and \eqref{eq:r-total-error-definition}. }
     \label{tab:r_estimates}
 \end{table}
 
 Alternatively  to   foreground  parameter maps, we also apply    clustering to partition the map of number of dust clouds along the line of sight, $\mathcal{N}_c$, presented in \citet{Panopoulou2020}. We perform the same parametric fitting as before on the patch derived from $\mathcal{N}_c$, and find similar level of residuals post-component separation for both \sosat and \lb cases when compared with ones obtained  from clusters  derived  with foreground parameter maps. Not only do we find that the partition obtained  is representative of the dust emission, but also that the patches inferred from $\mathcal{N}_c$ yield lower residuals in the estimates of the thermal dust temperature parameter, $T_d$.  
 
\noindent We further estimate the leakage in terms of $r$ also for this case of application  by means of the likelihood defined in  \eqref{eq:global-likelihood}  and we found $\hat{r} =(0.09\pm 0.26) \times 10^{-3}$ and $(0.04\pm 0.16) \times 10^{-3}$ respectively for \lb and SO-SAT. { The reason we obtain lower biases with respect to  the one reported in Table~\ref{tab:r_estimates} for the $f=0$ mismodelling case is mainly due to the fact that low frequency channels are  not accounted for in the component separation and we assume the synchrotron to be neglible at $\nu >90 $ GHz. Moreover,  we get a lower noise bias for \sosat with respect to \lb mainly because  to the fact that we do not account for the $1/f$ noise in our simulations for  both instruments.  }However, the low bias on  $r$ is a stricking indication that the $\mathcal{N}_c$ map can be a very promising tracer to probe the thermal dust emission in the range of $100-300 $ GHz. 
 Although
  \citet{Pelgrims2021}    showed  that the $\mathcal{N}_c$ maps  can be used as a  tracer for  frequency decorrelation in the dust polarization map, the results shown in this paper offer a further indication of their potential.

As already mentioned, this is not the first time that clustering methodologies are employed to   segment the sky into patches where to perform foreground cleaning (see  \citet{Grumitt_2020} where they used the mean-shift algorithm).  However, by comparing the maps of clusters  in Fig.~\ref{fig:clust_patches} with those in Fig.~2 of \citep[][]{Grumitt_2020}, we would like to address several reasons why  the two methods yield  very different   partitions. 

\noindent One immediate consideration is the fact that  mean-shift   and spectral clustering are based on  different algorithms that naturally might  yield  different results even when applied on  samples with fewer features than the ones in this context.
 
\noindent The clusters obtained in  \citet{Grumitt_2020} are larger than the ones shown in Fig.~\ref{fig:clust_patches}, although we also notice a  trend in estimating larger clusters  particularly for the $\beta_s$ clusters. This   makes  the method better suited for   low-resolution and  low-frequency channels where synchrotron emission dominates\footnote{An interesting application in  \citet{Grumitt_2020} was indeed the combination of the C-BASS data at 5 GHz with the \lb frequency channels.}, but it was essentially unable to track the finer structure of the dust parameters.

\noindent One of the major differences is   that we derive the clusters \emph{separately} for each spectral parameter, whereas in \citet{Grumitt_2020} the dust and synchrotron spectral indices $\beta_s$ and $\beta_d$  are combined together to derive a single cluster map used   to perform the parametric fit \citep[see Fig.~2 of][]{Grumitt_2020}. 

\noindent Furthermore, the features in \citet{Grumitt_2020} included the two spectral indices   together with the cartesian coordinates on the sphere $x,y,z$. On the contrary, our analysis does not rely on the  coordinates as features since  the adjacency is defined from the distance between pixels  (Section~\ref{subsec:heat_healpix}). We further account also for $T_d$ as an extra-parameter for the thermal dust emission, leading to a reduction in the foreground bias especially at high Galactic latitudes.

\noindent A novel feature of the technique presented in this work  is to perform the clustering by accounting for  the statistical  significance of the measurements  reported in the spectral parameter maps and their uncertainties. We weigh the pixel adjacency  in such a way that the significance   plays a non-negligible role in defining the pixel similarities especially in regions of high SNR. Moreover, we drastically reduce the dimensionality of our problem by performing the clustering on the eigenspace  spanned by the eigenvectors related to the smallest  256 eigenvalues. This avoids high-dimensionality problems occurring frequently in clustering algorithms with  large number of features (e.g. the number of pixels in a \texttt{nside=64} \textsc{HEALPix} map see Section 4 of \citet[]{Grumitt_2020}).

Finally, in Appendix~\ref{sec:gnilc}, we report also the clustering analysis performed on the GNILC $\beta_d$ and $T_d$ maps.  The morphologies are slightly different with respect to the ones observed in the clusters in Fig.~\ref{fig:clust_patches} as the spectral parameter features and the uncertainties in the two datasets are different.

\section{Conclusions} 
\label{sec:conclusions}

In this work, we aim at identifying regions in the sky within which the Galactic foreground emission can be assumed homogeneous.  We define a  partitioning of the full celestial sphere  by means of a novel technique based on a spectral clustering algorithm embedded in the $S^2$ manifold. 
  In contrast to previous applications of clustering for CMB studies, this method works directly on the celestial sphere instead of mapping points onto a Cartesian grid. It also takes into account the uncertainties in the clustering procedure. To our knowledge, this is  the first time  where  image segmentation has been performed on features defined on the  $S^2$ manifold.

Another key advancement of our algorithm is the use of an objective criterion for selecting the optimal partition parameters ($\alpha$ and $\delta$). Many \emph{off-the-shelf}  algorithms rely on subjective measures, such as visual inspection, to select a `reasonable' segmentation of the domain (e.g. the maximum pixel size parameter in mean-shift). The algorithm presented here borrows established metrics for partition measure evaluation in clustering problems to produce an optimal partition. We rely on the comparison of within- and between-cluster variances for this, while adding new modifications to incorporate the presence of uncertainties in the data.

We apply this clustering approach to evaluate Galactic contamination in CMB polarization studies.
 We firstly characterize the emission  of thermal dust and synchrotron available from the latest observations  of the spectral parameter maps (together with their uncertainties) to   divide the celestial sphere into regions with spectral clustering. 
 We then perform a parametric component separation for each region  simulating  the nominal frequency channels from two forthcoming CMB experiments observing from  the ground and from  space. 
 The optimal clustering partition  yields low residuals in the CMB polarization maps after component separation, even in the case where the Galactic emission is mis-modelled up to some extent.

  We further apply clustering to partition the map of number of \textsc{Hi}  clouds along the line of sight. Although this is one of the first applications using this kind of tracer, we find that clusters derived from the $\mathcal{N}_c$ map yield a residual bias as low as $r\sim 10^{-3}$ and that the dust spectral parameters are estimated on these regions up to $<10\%$ relative errors. This is a remarkable result since the  constraints on  $T_d$ from  far infrared data have been so far  complicated by the degeneracy with the spectral index $\beta_d$.  On the other hand, the use of ancillary data combined with the latest data-science  techniques  shows promising results in order  to improve the  modelling of  Galactic foregrounds.  
 
In conclusion, this technique could be applied on a wide range of contexts involving spherical images, e.g. Earth images, identification of Solar features,  wide  astronomical surveys and it can be easily extended to higher dimensional datasets, like  3D surveys or cosmological simulations.

\section*{Acknowledgements}


This research used resources of the National Energy Research Scientific Computing Center (NERSC), a U.S. Department of Energy Office of Science User Facility located at Lawrence Berkeley National Laboratory, operated under Contract No. DE-AC02-05CH11231.
Giuseppe Puglisi  acknowledges support by the COSMOS \& LiteBIRD networks of the Italian Space Agency.
The authors thank Clement Leloup, Carlo Baccigalupi for having read the paper thoroughly. 

Giuseppe Puglisi would like to thank: Jonathan Aumont, Mathieu Remazailles, Jens Chluba, Aditya Rotti, Susanna Azzoni, Leo Vacher, Hans Christian Eriksen, Nicoletta Krachmalnicoff  for useful comments and discussions.

Georgia Panopoulou acknowledges support for this work by NASA through the NASA Hubble Fellowship grant  \#HST-HF2-51444.001-A  awarded  by  the  Space Telescope Science  Institute,  which  is  operated  by  the Association of Universities for Research in Astronomy, Incorporated, under NASA contract NAS5-26555.

\section*{Data Availability}

The whole clustering analysis presented here has been collected into a  \texttt{python} package, ForeGround Clusters (\textsc{FGClusters} {\faGithub}\footnote{\url{https://github.com/giuspugl/fgcluster}}). 
 
The inputs used throughout this paper together with the outputs obtained with the Clustering technique have been made publicly available online\footnote{ \url{https://portal.nersc.gov/project/sobs/users/giuspugl/Clustering}} and in the Harvard Dataverse : 
\begin{itemize}
    \item $\mathcal{N}_c$ map : \url{https://doi.org/10.7910/DVN/XAMJ4X}; 
    \item PySM spectral parameters: \url{https://doi.org/10.7910/DVN/WJUEFA}; 
    \item GNILC dust parameters: \url{https://doi.org/10.7910/DVN/02SCHB}.
\end{itemize}




\bibliographystyle{mnras}
\bibliography{intro}




\appendix

\section{A geometric intuition on spectral image segmentation}
\label{appendixA}

The Laplace-Beltrami operator $\Delta_g$ is the generalisation of the Laplace operator $\Delta$ (divergence of the gradient) on a Riemannian manifold with metric $g$. In Local coordinates it can be written as:\[\Delta_g f =\frac{1}{\sqrt{|g|}} \partial_i\left( \sqrt{|g|} g^{ij}\partial_j f\right)\] The Laplace-Beltrami operator is a second order linear elliptic differential operator. It is the Hodge (connection) Laplacian acting on functions, it is by construction self-adjoint and positive-semidefinite.

The spectral properties of $\Delta_g$ on a compact differentiable manifold $M$ are well-known. The spectrum $\lambda_i$ is discrete, $\lambda_1=0$ is always an eigenvalue, all eigenvalues $0\leq\lambda_1...\leq\lambda_j\leq $ are positive. The eigenfunctions $\varphi_i$ of the Laplace-Beltrami operator on $M$ form a complete orthonormal basis of the space $L^2(M)$, i.e. any function in $L^2(M)$ can be written as a convergent series in $L^2(M)$ with real coefficients. This is the celebrated Sturm-Liouville decomposition of $L^2$ functions on smooth compact manifolds. The discussion of the eigenvalue problem on manifolds with boundary (with Dirichlet or Newmann boundary conditions) is often irrelevant in the discrete applications.

The Dirichlet energy functional of a function on $M$ is defined as\[E[f]=\frac{1}{2} \int_M ||\nabla f (x) ||^2 d\omega_g,\]\noindent where $\omega_g$ denotes the volume form over a Riemannian manifold. The Dirichlet energy measures the ``variability'' of $f$ on $M$.

The Dirichlet functional evaluated on the eigenfunctions $\varphi_i$ of $\Delta_g$ is monotonously increasing as the corresponding eigenvalues increase. The Stokes theorem applied on a function $f$ and a vector field $X$ implies that $-div$ and $\nabla$ are formally adjoint operators (see \cite{belkin}), i.e.:  \[\int_\mathcal{M} \langle X, \nabla f\rangle= -\int_\mathcal{M}  (\nabla \cdot  X )   f. \] \noindent So that: 
\begin{align*}
\int_\mathcal{M} \parallel  \nabla f(x)\parallel^2d\omega_g=\int_\mathcal{M}& \langle\nabla f, \nabla f\rangle\nonumber d\omega_g =\\
-\int_\mathcal{M}& (\nabla \cdot\nabla f)  f d\omega_g = -\int_\mathcal{M}  f(\Delta_g f) \nonumber d\omega_g.
\end{align*}

\noindent The former equation implies:  
\begin{align*}
E[\varphi_i]=\frac{1}{2} \int_M & ||\nabla \varphi_i (x) ||^2 d\omega_g=  \frac{1}{2} \int_M  \langle \nabla \varphi_i, \nabla \varphi_i \rangle d\omega_g=\\ \frac{1}{2} \int_M &\varphi_i (\Delta_g \varphi_i) d\omega_g= \frac{1}{2} \lambda_i
\end{align*}

\noindent For $f=\sum_i \alpha_i \varphi_i$ this expression linearly extends to \[E[f]=\frac{1}{2}\sum_i\alpha_i^2\lambda_i.\]\medbreak

The Laplace-Beltrami operator on $L^2(M)$ can be alternatively defined though the quadratic functional $E[f]$:\begin{equation}
E[f]=\langle \nabla f,\nabla f \rangle_{L^2(M)} = \langle \Delta_g f, f\rangle_{L^2(M)}.
\label{dirichlet}
\end{equation}Interestingly, $\nabla E[f]=\Delta_g f$ i.e. the Laplace-Beltrami operator  is the variational derivative of Dirichlet energy in $L^2(M)$.\medbreak

\begin{Proposition}
The function that minimises the Dirichlet energy and is orthogonal in $L^2(M)$ to the space spanned by $\varphi_0,...,\varphi_i$ is $\varphi_{i+1}$.
\end{Proposition}

\subsection{Fredholm theory and discrete differential operators}

Digitally produced images resemble smooth functions sampled discretely  over regular domains (a grid of pixels),  the above construction needs to be replicated in a discrete form. For some general ideas about the discretisation of differential operators acting on sections of tensor bundles, we refer the reader to \cite{mihaylov}). Manifold learning techniques that include consistent discretisations of key elements of Riemannian geometry (vector fields, connections) have been described in \cite{singer}. Recently this approach has been further developed in \cite{berry2} including differential forms, Laplace-De Rham operators etc.

Let us consider the following differential equation:\begin{equation}\mathcal{D} f= g, 
\label{defGreen}\end{equation}
where $\mathcal D:L^2(M)\longrightarrow L^2(M)$ is a linear differential operator. 
The Green's function of a differential operator is a special integral kernel related to the Dirac's delta.  \[\mathcal{D} G(x,y)=\delta(x-y)\] As an application of Fredholm's theory, the solution of Equation~\eqref{defGreen} can be written in the following equivalent integral form:\[f(x)=\int_M G(x,y)g(y) d\omega .\] For operators with discrete spectrum and a complete orthonormal basis of eigenfunctions $\{ \varphi_i \}$ the Green's function can be expressed:\begin{equation}
G(x,y)=\sum_i\frac{1}{\lambda_i}\varphi_i(x)\varphi_i(y).
\label{greentensor}
\end{equation}

If we consider a sampling of $n$ points $x_i \in M$ and choose a conventional order for the sampled points, a finite sampling of a function $f$ can be represented by a real n-dimensional vector. Denote by $G(x,y)$ the Green's function of a linear differential operator $\mathcal{D}$ on $M$. In line with \cite{mihaylov}, we call a \emph{coarse discretisation} of $\mathcal{D}$, the linear map $D:\mathbb{R}^n\longrightarrow \mathbb{R}^n$ defined by:\begin{equation}D G(x_i,x_j)=\delta_{ij},\label{inversa} \end{equation} \noindent where $\delta_{ij}$ denotes Kronecker's symbol. With this definition, the continuous eigenfuction problem $\mathcal{D}f=\lambda f$ is discretised into \begin{equation}\sum_{i}G(x_i,x_j)f(x_j)=\mu f(x_i),
\label{inverse}
\end{equation} The Green’s function, measures the way in which two points are geometrically related.  In the above problem, the contribution of the sampling $f(x_j)$ to the computation of $f(x_i)$ can be further localised by introducing an adjacency matrix $W_{ij}$ which provides the sampling $x_i\in M$ with a graph structure. The elements of $W_{ij}$ are equal to 1 or 0 based on different criteria depending on the problem, a distance-based cutoff, a fixed number of neighbouring points (pixels), etc. Equation~\eqref{inverse} becomes:\begin{equation}
Df(x_i)=\sum_{jk}G(x_i,x_j)W_{jk}f(x_k)=\mu f(x_i).
\label{elements}
\end{equation} Self-adjoint operators give rise to symmetric discretisations and the discrete version of Equation~\eqref{greentensor} is:\[G(x_m,x_n)=\sum_i\frac{1}{\lambda_i}\varphi_i(x_m)\varphi_i(x_n).\]This equation and Equation~\eqref{inversa} mean that $D$ is the generalised inverse of $G$ (or $GW$). By construction, if we increase the density of the sampling as $n\longrightarrow \infty$, the coarse discretisation converges to the differential operator.

The coarse discretisation of a known differential operator on a given manifold is a conceptually straightforward process and it differs from the typical manifold learning problem of approximating a linear differential operator $\mathcal{D}$ on a point cloud sampled from a priori unknown sub-manifold of $\mathbb{R}^n$. We call such a linear map the sample discretisation of $\mathcal{D}$.

The elements used for building a sample discretisation are the same as in Equation~\eqref{elements} i.e. an affinity graph and a kernel function $K(x_i,x_j)$ that satisfy certain symmetry and regularity assumptions (see \cite{singer, berry}). In the manifold learning setup the kernel $K(x_i,x_j)$ reproduces the expression of the (unknown) Green's function of the manifold in terms of the coordinates of the points $x_i$ in the space of measured variables. A first approximation attempt regarding the expression of the kernel can be found between the Green's function of the same operator on $\mathbb R^n$ computed with the embedding coordinates $x_i$. 

This estimate must be corrected by bias terms that include for example the curvature on $M$, the sample variance etc. Intuitively, the bias of the geodesic distance on $M$ with respect to the Euclidean distance in the ambient space is measured by the curvature. 

Given a sample discretisation of a differential operator, two types of convergence problems arise in the limit of infinite sampling:  
\begin{itemize}
\item point convergence $D\varphi(x_i) \longrightarrow  \mathcal{D} \varphi (x)$
\item spectral - eigenvectors of $D$ converge to eigenfunctions $\mathcal{D}$.
\end{itemize}

Spectral convergence is a stronger condition. Proving the convergence of a discretised operator is a non-trivial analytical problem that includes the curvature of the manifold,  the density of sampling, normalization choices etc. Remarkable results in this direction have been achieved in \cite{belkin, singer}. \subsection{Discrete Laplacian} 

For a point cloud in $\mathbb{R}^n$ we define a graph with adjacency matrix $A_{ij}$ with weights $K(x_i,x_j)$ and compute the (weighted) diagonal degree matrix $D_i=\sum_j A_{ij}$. There are several definitions of Laplace operators on graphs that use these elements. Typically the adopted expressions for $K(x_i,x_j)$ are inspired by the Laplacian Green's function in the ambient space of the point cloud. For example the Green's functions of the Laplacian in $\mathbb{R}^2$ and $\mathbb{R}^3$ are well known: \[G_{\mathbb{R}^2}(x,y)=\frac{1}{2\pi}\ln (|x-y|), \,\,\,\,\,\,\,\,\,\, G_{\mathbb{R}^3}(x,y)=\frac{1}{4\pi|x-y|} .\] \noindent \citet{berry} studied more  general kernel constructions that take into account non-uniform sampling densities. 
 
The most popular construction is the   \textbf{graph Laplacian}\begin{equation} L:=D-A.\label{graphL}
\end{equation}

This definition arises from the Newton's cooling law on a graph. The heat transferred between connected nodes $i$ and $j$ is proportional to the difference $f(x_i)-f(x_j)$ and a heat capacity coefficient $k$, i.e. \begin{align*}
    \frac{df(x_i)}{dt}=&-k\sum_j A_{ij}(f(x_i)-f(x_j))=\\
    &-k\left( f(x_i) \sum_j A_{ij} - \sum_j A_{ij} f(x_j)\right)
\end{align*}This is further motivation for the degree matrix $D$ in Equation~\eqref{graphL}.\medbreak

 \noindent The \textbf{random walk Laplacian} is defined as \begin{equation}L_{rw}:=D^{-1}L=I-D^{-1}A\label{RWL}\end{equation} and is related to Markov processes on graphs. The matrix $L_{rw}$ is not symmetric, but row-stochastic (satisfies the Markov property). This definition has been further developed in the theory of diffusion maps (see \cite{coifman,singer} etc).\medbreak

\noindent  The \textbf{symmetric random walk Laplacian} is:\begin{equation}L_{sym}:=D^{-\frac{1}{2}}L D^{-\frac{1}{2}}=I-D^{-\frac{1}{2}}A D^{-\frac{1}{2}}.\label{normL}
\end{equation}
\noindent The relations between these definitions and the transformations that map their spectral structures respectively are well-known. 

Remarkably,  $L_{sym}$ matrix approximates the Laplace-Beltrami operator   both  in pointwise and spectral sense (see  \cite{belkin} and \cite{singer}). More precisely for $n\longrightarrow \infty$,\[L f(x_i)\approx \Delta f(x)+O\left(\frac{||\nabla f(x)||}{\sqrt{n}\epsilon^{1/2+dim(M)/4}}\right),\]
\noindent where $\epsilon$ is a parameter (see \cite{berry}). Moreover, in the    case of uniform sampling, \cite{belkin}   have shown that the eigenvectors of the graph Laplacian converge to the eigenfunctions of the Laplace-Beltrami operator on the manifold. 

Constructions of discrete Laplacian operators from non-uniform distributions of manifold have been developed. Sample density can be absorbed in the definition of the kernel itself or a ``right normalisation'' is introduced \cite{berry}. These constructions are less relevant for the purposes of image segmentation where images are usually sampled on regular girds of pixels. 

The geometric relation between kernels and Riemannian metrics has been investigated in \cite{berry} for uniform and nonuniform sampling. The expression of the Laplace-Beltrami operator changes if the Riemannian metric on $M$ varies in a family of metrics, which includes the one induced by the embedding $M\hookrightarrow \mathbb{R}^n$. These changes are captured by modifications in the form of the Green's function of the operator and subsequently in its discretisation. Every symmetric local kernel with exponential decay corresponds to a Laplacian operator in a Riemannian geometry and vice versa. 

Once a discrete Laplacian is provided, the discrete Dirichlet energy functional of Equation~\eqref{dirichlet} applied on a sampling $f(x_i)$ becomes: \[E[f(x_i)]=\frac{1}{2}\langle f(x_i),L f(x_i)\rangle.\] The construction of low-energy embedding of a graph in $\mathbb{R}^m$ with $m$  lower or equal to the number of sampled points is directly applied in the discrete case and is used in spectral image segmentation.

\subsection{Spectral segmentation by means of the heat propagator}  

The geometric affinity/similarity between points on a given manifold can be measured by a diffusion process. Eq.\eqref{eq:heat_diff} is the heat diffusion differential equation on a smooth manifold.

The Laplace-Beltrami-based construction described above can be replicated by by means of a closely related differential operator $e^{-t\Delta_g}:L^2(M)\longrightarrow L^2(M)$ called (for $t>0$) the \emph{ heat propagator}. Fredholm theory in this case is applied through a Green's function called the \emph{heat kernel} $G(x,y,t)$. Similarly to the Schr\"odinger equation, the heat kernel is a special solution of\[\Delta_g G(\mathbf{x},\mathbf{y},t) =\partial_t G(\mathbf{x},\mathbf{y},t),
\,\,\,\text{ such that }\,\,\,  
G(\mathbf{x},\mathbf{y},t)=e ^{t\Delta_g} \delta (\mathbf{x}-\mathbf{y} ).\] The heat kernel is used to solve in integral form the diffusion equation with a source term $f(x)$ and with boundary conditions:\begin{equation}
u(x,t)= e^{-t\Delta_g} f(x)=\int_M G(x,y,t)f(y)d\omega_g
\label{eq:DiffusionSolution}
\end{equation}\noindent The heat propagator  inherits from the Laplace-Beltrami operator the  properties of being  a self-adjoint,  positive-definite and compact. 
Its eigenfunctions form a complete orthonormal basis of $L^2(M)$. The eigenvalue problem for the heat propagator becomes: \begin{equation}
e^{-t\Delta_g}\varphi_i=\beta_i^t \varphi_i
\label{eigenexp}
\end{equation} Let us set $\lambda_i:=-ln(\beta_i)$. By construction $e^{-t\Delta_g}\varphi_i$ satisfies the heat diffusion equation for all $i$. Substituting Equation~\ref{eigenexp} we get:\[ 0=L[e^{-t\Delta_g}\varphi_i]= e^{-\lambda_i t}(\Delta_g \varphi_i-\lambda_i\varphi_i).\]  So $\varphi_i$ are precisely the eigenfunctions of the Laplace-Beltrami operator with eigenvalues $\lambda_i$. As a consequence, the relevant properties that enable image segmentation through the Laplace-Beltrami operator (eigenfunctions define a low Dirichlet energy embedding of $M$ in $\mathbb{R}^n$) can be directly extended to the heat propagator. In particular
\begin{equation}
   G(x,y,t)=\sum_i e^{-\lambda_i t} \varphi_i(x)\varphi_i(y). 
\label{eq:tensor_eigenf}
\end{equation}
 For instance, the heat kernel in $\mathbb{R}^3$ is: \[G(\mathbf{x},\mathbf{y},t)_{\mathbb{R}^3} = \left(\frac{1}{4\pi t}\right)^{3/2} \exp\left(  - \frac{\parallel \mathbf{x}-\mathbf{y} \parallel ^2}{4t}\right)\,\,\, \text{,}\,\,\,  G(\mathbf{x},\mathbf{y},0) = \delta (\mathbf{x}-\mathbf{y}) \] An argument provided in \cite{belkin} derives Equation~\eqref{graphL} by discretising the heat propagator. By substituting in \ref{eq:DiffusionSolution} in \ref{eq:heat_diff} we get:

\begin{align*}
\Delta_g f(x)=\Delta_g u(x,0)=&\frac{\partial }{\partial t} \left(\int_M G(x,y,t)f(y)d\omega_g\right)_{t=0} \nonumber \\
\approx& -\frac{1}{t}\left( f(x)-\int_M G(x,y,t)f(y)d\omega_g\right) \\
\approx&-\frac{1}{t}\left( f(x_i)-\sum_j G(x,y,t)f(x_j)\right). 
\end{align*}

The overall coefficient $1/t$ does not affect the spectrum of the expression in the brackets. This argument is in line with the above discussion on the heat propagator and justifies further the diagonal term in the discrete definitions of the discrete Laplacian based on diffusion arguments. 

Pointwise and spectral convergent discretizations of the heat propagator have been developed in the manifold learning theory of diffusion maps ( \cite{coifman,singer,berry}). Given the functional form form of the heat kernel in $\mathbb{R}^n$ this class of discrete constructions of the Laplace-Beltrami operator predominantly exploit global and local Gaussian kernels.

\subsection{Heat-kernel in $S^2$ }
\label{app:heat_kernel}

As discussed in Sect.\ref{sec:spectral}, we choose  the adjacency weights to be the ones  derived from the functional form of the integral kernel of the Laplace-Beltrami operator in $S^2$. 
In this section, we  derive the functional form in eq.~\eqref{eq:adj_heat}. 

First,  we write  the heat equation in spherical coordinates: 
\begin{equation}
    (\partial_t - \Delta _g)u(x,t) = L[u(x,t)]=\rho (x,t) , 
    \label{eq:heat_diff}
\end{equation}
with $\Delta _g$ being the Laplace-Beltrami operator in $S^2$, $u$ the unknown function describing the heat diffusion, $\rho$ the source term,  $x$ and $t$  respectively the spatial and time variables.    We can express $\Delta _g$ in spherical coordinates (assuming a radius $r=1$):
\begin{equation}
 \Delta \equiv \frac{1}{  \sin \theta }\frac{\partial}{\partial {\theta} }  \left(\sin \theta \frac{\partial}{\partial {\theta} } \right) + \frac{1}{ \sin^2 \theta }\frac{\partial^2}{\partial{\phi^2 } }.
\end{equation}
Expressed in this way, we can easily recognize that the Laplace-Beltrami operator  corresponds to  $ 
  -\hat{L}^2$,  i.e. the square of the  \emph{orbital angular  momentum}  operator  in quantum mechanics,  whose eigenfunctions are the \emph{spherical harmonics}, $Y_{\ell m}$, and whose eigenvalues are $-\ell(\ell+1)$ ($\ell \in \mathbb{N}$).  We further notice that the heat propagator operator, $\exp{(- \hat{L}^2 t )}$ shares the same eigenfunctions and eigenvalues as $- \hat{L}^2$.
In Appendix~\ref{appendixA}, we show how the integral kernel of the heat equation is related to the eigenfunctions of $\exp{(- \hat{L}^2 t )}$ operator (see ~\ref{eq:tensor_eigenf}), we can thus derive the   kernel  in spherical coordinates\footnote{Assuming as an initial condition to the  source term, $\rho$  in \eqref{eq:heat_diff} to be  a Dirac's delta distribution, i.e. $\rho(x,0) = \delta (x) .
$}:

\begin{align}
G(\mathbf{x},\mathbf{y},t)=& \sum_{\ell =0} ^{+\infty } e^{-t \ell(\ell+1)}  \sum _{m=-\ell} ^{+\ell} Y_{\ell m}(\mathbf{x}) Y_{\ell m}^* (\mathbf{y}) \nonumber  \\ 
=& \sum_{\ell =0} ^{+\infty } \frac{2\ell +1}{4\pi}   e^{-t \ell(\ell+1)} \mathcal{P}_{\ell}(\Theta_{\mathbf{xy}}), 
\label{eq:heat_kernel}
\end{align} 
where we exploit in the last equality  the properties of spherical harmonics, see further details in Appendix~\ref{appendixA}.  In particular, the last equality is obtained  by expressing the $Y_{\ell m}$ as a function of  \emph{Legendre polynomials}, $ \mathcal{P}_{\ell}$, evaluated across the \emph{cosine matrix},  defined as : 
\begin{equation}
    \Theta_{\mathbf{xy}} \equiv \cos\theta_{\mathbf{xy}} = \frac{\mathbf{x} \cdot\mathbf{y}} {\parallel \mathbf{x}\parallel \parallel \mathbf{y}\parallel }.
    \label{eq:theta}
\end{equation} 

 We devote Sect.~\ref{sec:spectral} to describe how to derive the  adjacency weights  from eq.~\eqref{eq:heat_kernel}, so that we can construct the Laplacian matrix as in eq.~\eqref{eq:laplacian_sym} and estimate the eigenpairs to perform the spectral clustering.

\subsection{Laplacian eigenfunctions and the Ricci curvature}

There are two conceptually distinct ways in which eigenfunctions of the Laplace-Beltrami operator and their discrete analogues are exploited for image processing. \medbreak

- An image is interpreted as a function on a regular domain (typically a rectangle in $\mathbb{R}^2$). This function is decomposed with respect to the special basis of eigenfunctions. This decomposition provides a useful characterisation of an image, allows highlighting or filtering global components with specific Dirichlet energy i.e. spatial ``frequency''. As an elementary example the coefficient of $\varphi_0$ captures an overall intensity. This approach, similar to spectral shape analysis, has been adopted in \cite{reuter} and \cite{rustamov} and is not typically used for segmentation purposes.\medbreak

- An image itself is considered as smooth surface (real 2-dimensional manifold) sampled in a discrete set of pixels. The geodesic distance between points on the manifold combines the distance between pixels with the the ``vertical deformation'' that encodes the image.  Spectral geometry is a field of differential geometry that exploits the spectral structure of Laplace-Beltrami operator to describe relevant geometric and topological properties of manifolds. The spectral structure of the discrete Laplacian approximates the spectral structure of the Laplace-Beltrami operator. This is the manifold learning approach to spectral segmentation.

The application of ideas from geometry and theory of partial differential equations to image segmentation is an active field of research. Variational calculus based on a variety of energy functionals (Mumford-Shah, Chan-Vese, elastic energy with additional forcing terms and well potentials) and dynamics on different time scales have been developed in \cite{Bertozzi1, Bertozzi2, Bertozzi3}. The deep geometric nature of the spectral clustering methods is yet to be investigated. In fact the formulation of spectral image segmentation in the smooth case, means that points in plateau-ing regions of the manifold are close to level sets of the eigenfunctions of the Laplace-Beltrami operator in a suitable Dirichlet energy band. Relevant properties of the nodal (and more general level) sets of the eigenfunctions of $\Delta_g$ have been studied in special geometric cases (constant curvature or curvature bounded from below). Less is know on how the level sets of eigenfunctions capture the geometry of a manifold in the general case. 

Equivalently, we expect that eigenfunctions of the Laplace-Beltrami operator with given Dirichlet energy are characterised by small variations on plateau-ing regions of the manifold. The key question is how can we characterise platteau-ing regions. 

Our hypothesis is that these are the regions in which the Ricci curvature of the manifold is low (either positive or negative). A version of the celebrated Bochner's formula establishes a direct relation between the Ricci's curvature tensor, the Laplace-Beltrami operator and the norm of the Hessian of a smooth function $\varphi$ on a Riemannian manifold $M$ with metric $g$: \[\frac{1}{2}\Delta_g |\nabla \varphi|^2 = g(\Delta_g \nabla \varphi,\nabla \varphi)+|H(\varphi)|^2+Ric(\nabla \varphi,\nabla\varphi).\] \noindent For eigenfunctions $\Delta_g\varphi=\lambda \varphi$ the Bochner's formula becomes: \[Ric(\nabla \varphi, \nabla \varphi) = \frac{1}{2}\Delta_g |\nabla \varphi|^2 - \lambda |\nabla\varphi|^2-|H(\varphi)|^2\] For smooth surfaces (images) this equation means that the sectional curvature in the direction of the maximum variation of the eigenfunction depends directly on the magnitude of the gradient of the eigenfunction and the eigenvalue.  

The above considerations suggest the existence of a deep geometric reason for the fact that spectral image segmentation is very efficient in detecting objects with complicated shapes in unsupervised setup. Understanding spectral image segmentation in the context of spectral geometry is a promising field of future research.

\section{Clustering applied on \emph{Planck}-GNILC dust maps  } \label{sec:gnilc}
 \begin{figure*}
      \centering
      \includegraphics[scale=.5]{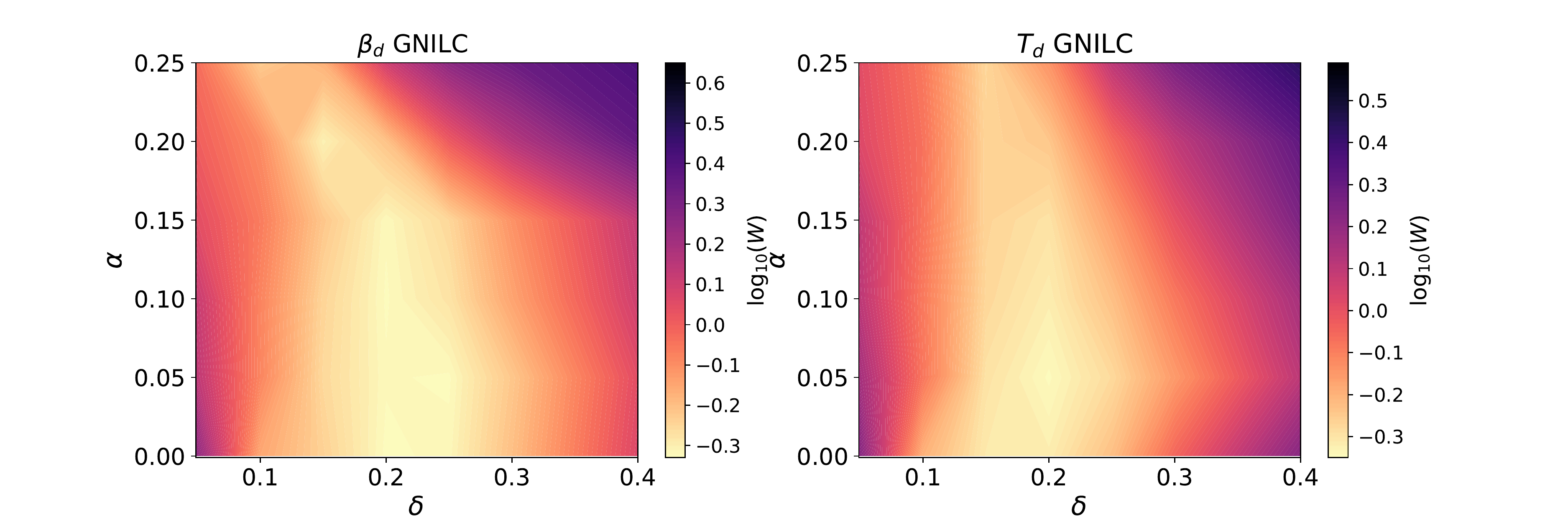}
      \caption{Variance surfaces estimated for  several choices of $\alpha$ and $\delta$  for (left)  $\beta_d$, (right) $T_d$. Notice that regions of optimality corresponds to local minima (lighter regions) in the surfaces.  }
      \label{fig:variance_gnilc}
  \end{figure*}
   \begin{figure*}
     \centering
     \includegraphics[scale=.25, trim=0 0 0  4cm, clip=true ]{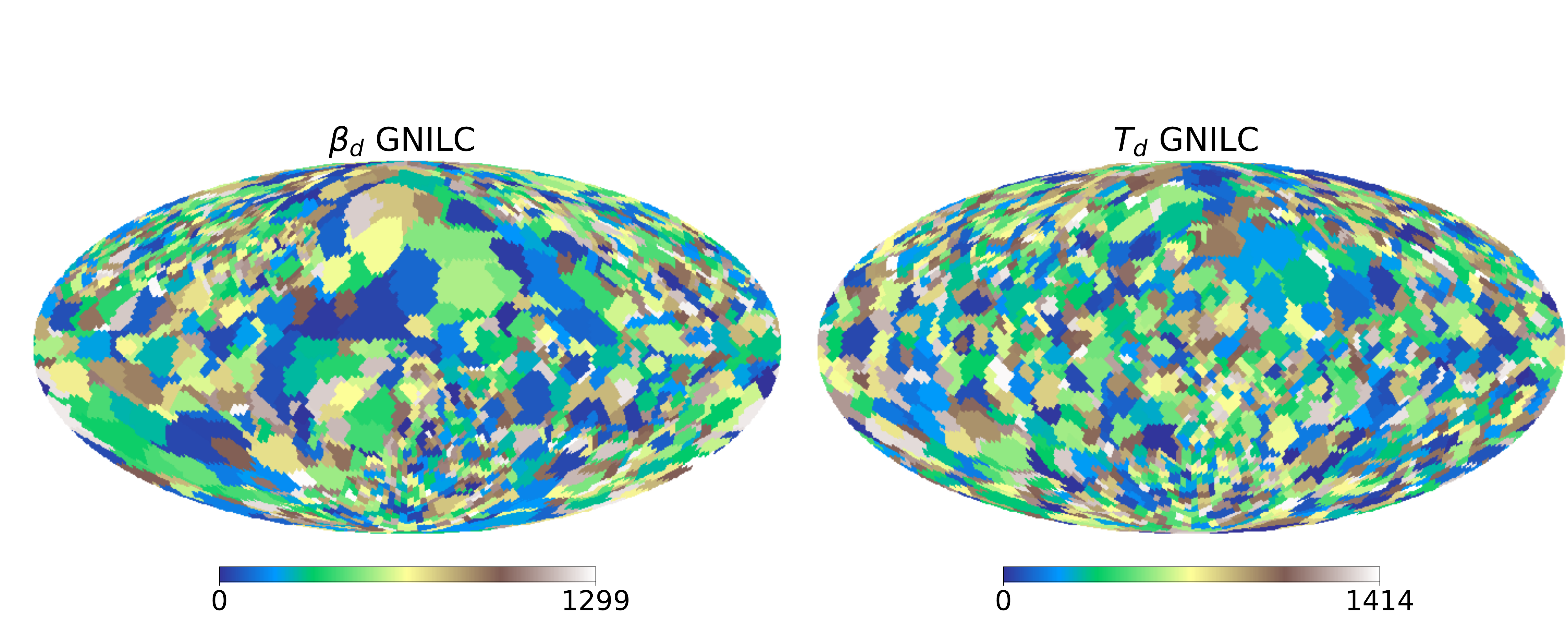}
     \caption{Cluster maps of (left) $\beta_d$ and (right) $T_d$ GNILC maps.}
     \label{fig:clus_gnilc}
 \end{figure*}

 \begin{figure*}
     \centering
     \includegraphics[scale=.364] {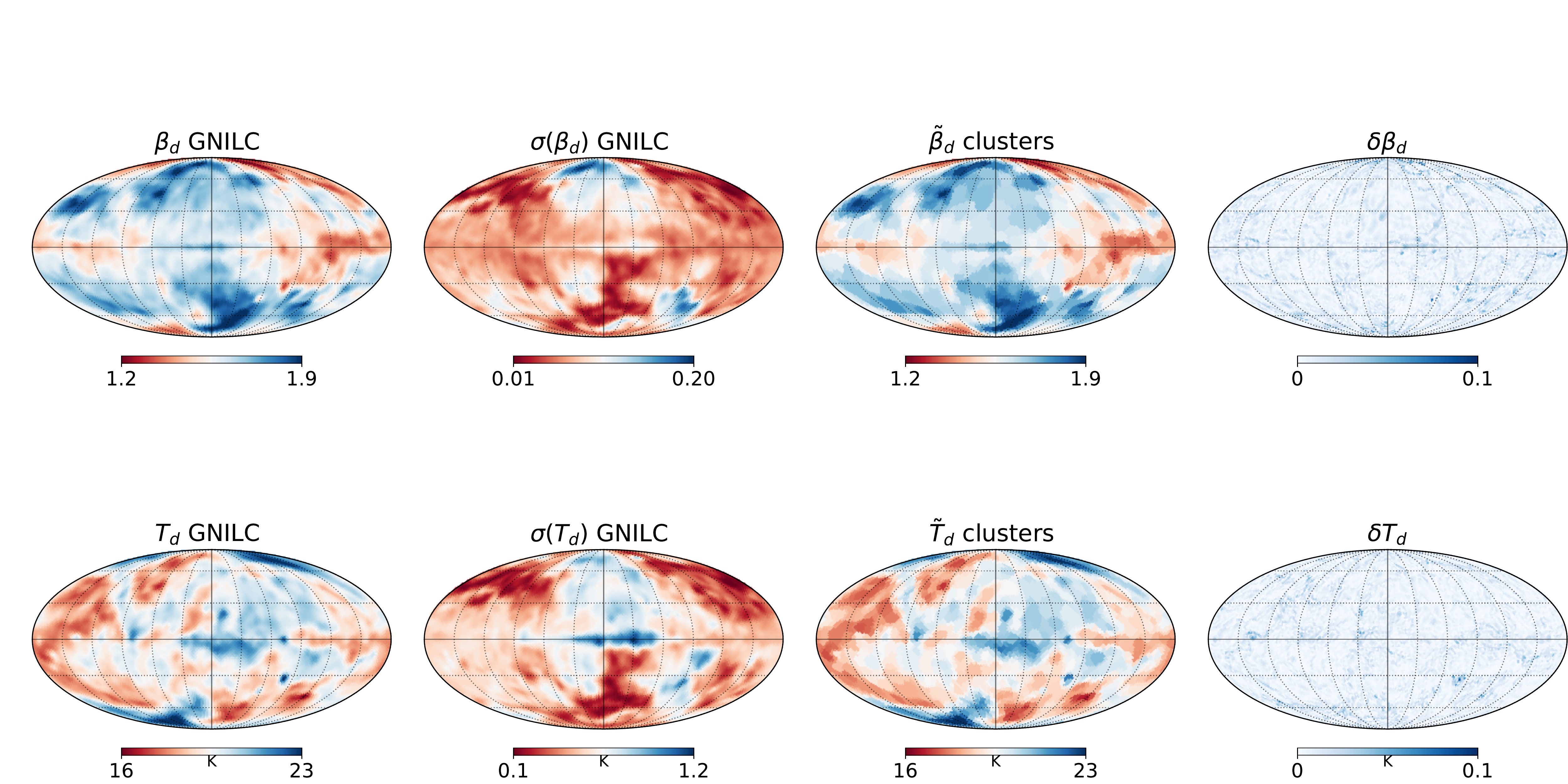}
     \caption{(first and second column) Dust  parameter and uncertainty  maps  from \planck -GNILC. (third column)   Dust spectral parameter maps evaluated by taking  the median value of the input GNILC maps on each region defined by  spectral clustering procedure. (fourth column) Relative errors estimated by taking the difference of GNILC   maps with the binned ones.   }
     \label{fig:maps_gnilc}
 \end{figure*}

  \citet{gnilc2016} released  a set of maps obtained by separating the Galactic thermal dust emission from the cosmic infrared background (CIB) anisotropies. They  implemented  a   tailored component-separation method, the  Generalized Needlet Internal Linear Combination (GNILC) method, that adopts  spatial information (e.g. the angular power spectra) to separate  the Galactic dust emission and CIB anisotropies. 
  We thus considered the temperature $T_d$ and spectral index $\beta_d$ maps of thermal dust  and the uncertainties accompanying these maps to partition the sky into multiple domains accounting both for the geometrical similarities and the uncertainties.  
  
  Given the fact that we are interested to derive patches at the $\sim$ degree scales, we firstly reduce the resolution of the GNILC maps (released at \texttt{nside=2048} and $5'$ resolutions),  to a coarser beam, $110 '$ and to a lower  pixel resolution,  \texttt{nside=32}. This also makes the Laplacian matrix computation faster with less memory requirements.

  We then performed the clustering following the Algorithm \ref{alg:spectral_clus} separately for $T_d$ and for $\beta_d$. In Fig.\ref{fig:variance_gnilc},  we show the variances estimated from the GNILC maps in an $\alpha-\delta$ grid. We note the presence of a minimum variance vertical stripe at around $\delta\sim 0.2$ and for $\alpha\leq 0.2$.  Notice that high variance regions can be identified in the upper right and lower left corners of the image related respectively to over and under partition regimes. Moreover,    ranges of optimality both for $\beta_d$ and $T_d$ present  similar variance contours.  
 Since  the optimization is performed independently for the two dust parameters,  we interpret this result as an indication that  the methodology might be optimizing  for similar Galactic scale features  in both $\beta_d$ and $T_d$ maps.  

 We find the optimal number of clusters to be: 1300 and 1415 respectively for $\beta_d$ and $T_d$ shown in Fig.~\ref{fig:clus_gnilc}, corresponding to the parameter values of $\alpha=0.15,\delta=0.20$.

 Finally, we want to assess how well the optimal regions track the observed features in the GNILC maps. We thus estimate within each cluster patch, the median value of the pixels identified in a single cluster and build a \emph{binned} map for $T_d$ and $\beta_d$, labeled as $\tilde{T}_d$ and $\tilde{\beta}_d$. We show in Fig.\ref{fig:maps_gnilc} the maps from GNILC and the binned maps. To quantify the error of this procedure, we estimate the  relative   difference of the input and binned maps as  in eq.~\eqref{eq:rel_res}. 
 
  Relative error maps  are shown in the right-most column of Fig.\ref{fig:maps_gnilc}.  We note that the errors are  lower than the $10 \%$.

\section{Defining the optimality range}\label{app:optimality}
\begin{figure*}
    \centering
    \includegraphics[width=1.95\columnwidth, trim=1cm 0 12.5cm 3cm , clip=true] {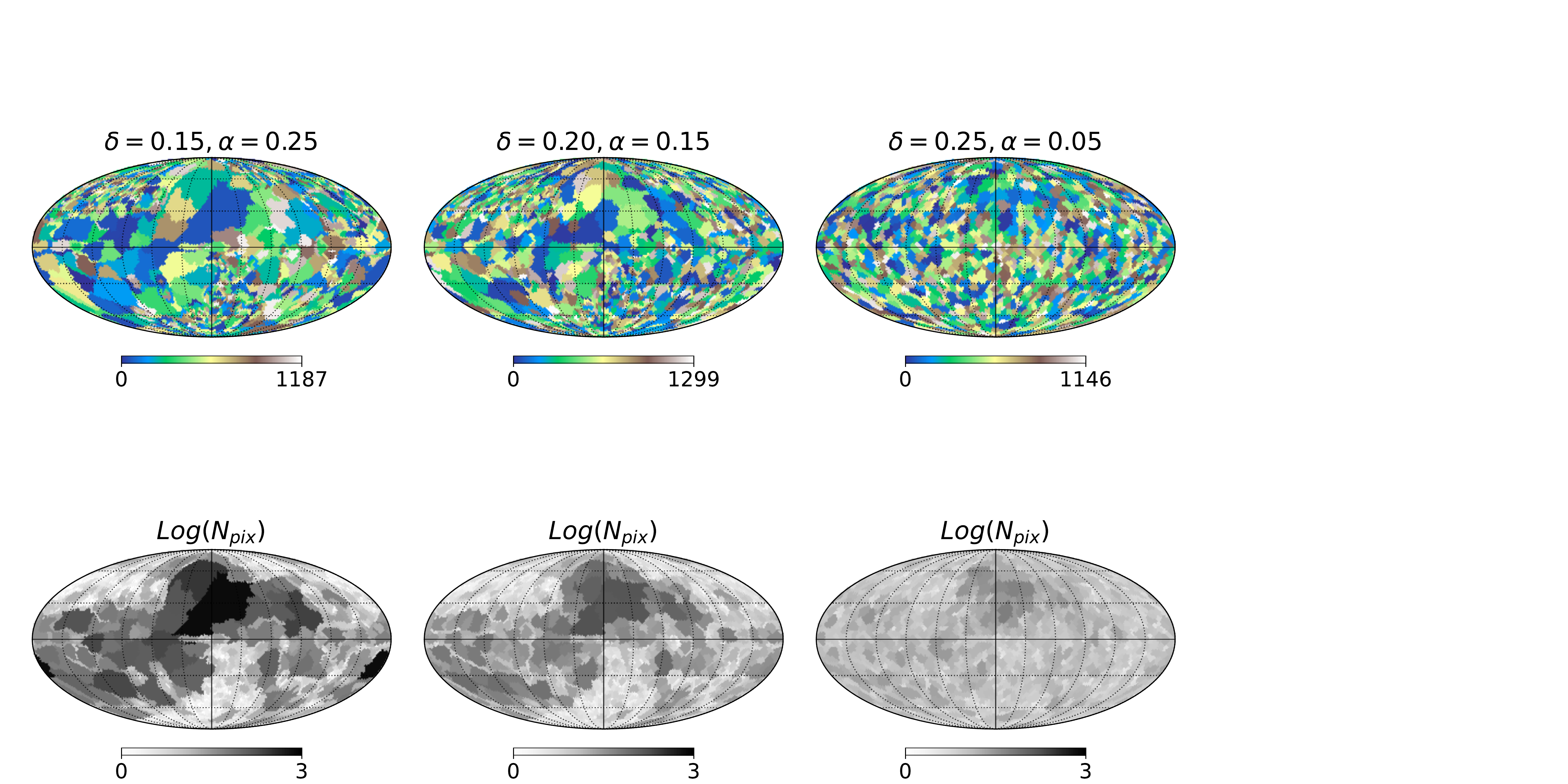}
    \caption{(top) Cluster patches for $\beta_d$   GNILC maps, in regions of optimality as shown in Fig.~\ref{fig:variance_gnilc}, for several choices of $\alpha$ and $\delta$. (bottom)  Maps showing the logarithm of the number of pixels within each  cluster shown in the top row.   }
    \label{fig:size_clust}
\end{figure*}
From the variances shown in Figures ~\ref{fig:variance_gnilc}, \ref{fig:fg_var}   and  \ref{fig:clus_variance}, we notice that the range of optimality corresponds  to a narrow range  in $\delta$  but seldom involves a broader range in the $\alpha $  parameter. 

In the top row of  Fig.~\ref{fig:size_clust}, we show several cluster configurations chosen in the range of optimality from the variance in Fig.~\ref{fig:variance_gnilc}(left).  We quantify the size of the cluster by estimating how many pixels belong to each cluster. The logarithm of the number of pixels per cluster is shown in the bottom row of Fig.~\ref{fig:size_clust}.  Values of $\alpha<0.1$, lead to a very homogeneous and isotropic partition which is intuitively  expected as we  partition the celestial sphere given the properties of the symmetric Laplacian adjacency (eq.~\eqref{eq:adj_heat}).  On the other hand,  increasingly larger values of $\alpha>0.2$  make the cluster morphology   more and more dictated by the SNR content of the  features, resulting in inhomogenuous sizes and anisotropic shapes in the vicinity of the Galactic plane as the manifold adjacency    gets more and more distorted (see eq.~\eqref{eq:deform}. 
We find that intermediate values for $\alpha$ in the range around $0.1 \div 0.2$  result in a good trade off between these two extreme cases.
 
 \section{Estimation of uncertainties for the $\mathcal{N}_c$ map}
 \label{appendix:Nc}
 
 As described in Section \ref{sec:nc}, we apply our clustering algorithm to partition the sky in regions of near-constant number of clouds per line of sight, as determined by the analysis of HI data in \citet{Panopoulou2020}. A critical component of the clustering analysis is its ability to handle measurement uncertainties. The public map of $\mathcal{N}_c$ does not contain uncertainties and so we repeat steps of the original analysis to obtain estimates of the per-pixel uncertainties in the map. 
 
 We compute the uncertainty on $\mathcal{N}_c$ by taking into account the two sources of error that enter in the calculation of $\mathcal{N}_c$: 
\begin{enumerate}
    \item the uncertainty on the number of velocity  peaks identified in the HI spectrum, 
    \item the uncertainty on the column density of each cloud.
\end{enumerate}

As explained in \citet[Appendix B1]{Panopoulou2020}, both sources of uncertainty are driven by the choice of the velocity kernel size (a parameter that implements smoothing in velocity space, also referred to as the bandwidth).   The choice of bandwidth presents a trade-off between resolving power (ability to distinguish nearby peaks in velocity) and fidelity (avoiding spurious peak detections). Furthermore, the kernel size also alters the shape of the probability distribution, including the position of extrema and saddle points,  used to define the range of each peak. 

We therefore  estimate the uncertainties  by considering  maps of $\mathcal{N}_c$ with  three different choices  of  bandwidth: namely 3, 4   and   5 velocity channels, indicated  respectively as $\mathcal{N}_c^{\mathrm{bw3}}$, $\mathcal{N}_c^{\mathrm{bw4}}$, $\mathcal{N}_c^{\mathrm{bw5}}$, with $\mathcal{N}_c\equiv \mathcal{N}_c^{\mathrm{bw4}}$ being the default value in \citet{Panopoulou2020}.  We evaluated the relative differences of $\mathcal{N}_c$ between different bandwidth runs as:
\begin{displaymath}
\Delta_{rel}^{i, 4}=  (\mathcal{N}_c^{\mathrm{bw\,i}} - \mathcal{N}_c^{\mathrm{bw4}}) / \mathcal{N}_c^{\mathrm{bw4}}. 
\end{displaymath}
These differences can be used to quantify the systematic uncertainty of the cloud identification method, and hence $\mathcal{N}_c$. 

\begin{figure}
     \centering
     \includegraphics[scale=1] {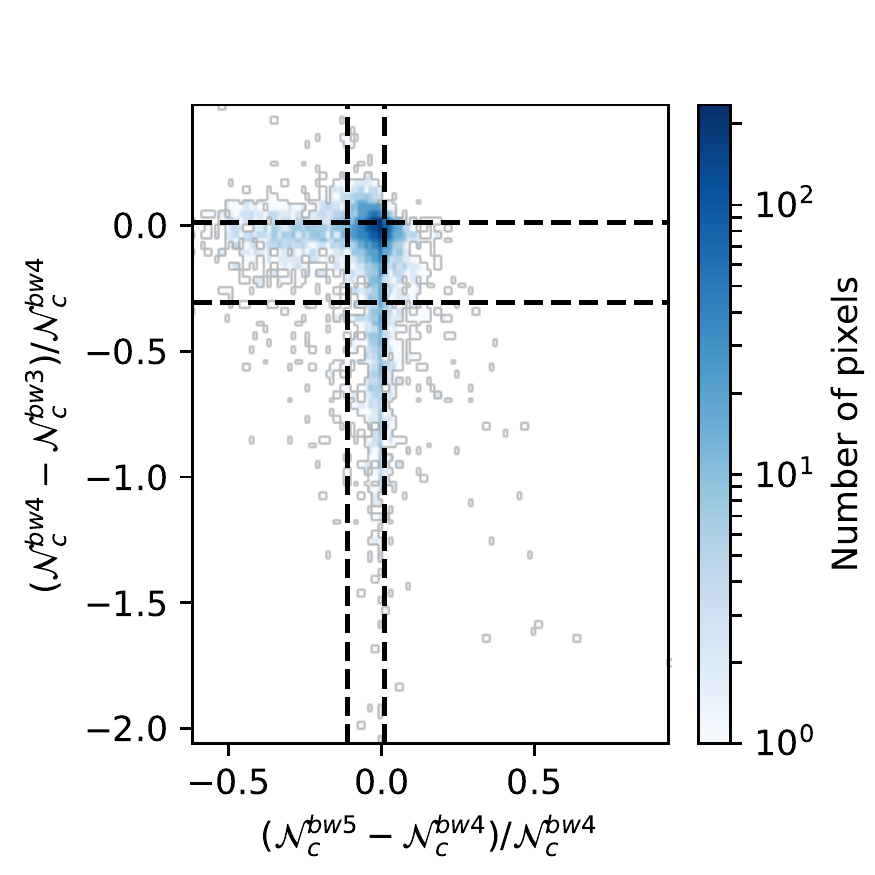}
     \caption{Distribution of asymmetric uncertainties on $\mathcal{N}_c$ at \texttt{nside=32}, using the relative difference between $\mathcal{N}_c$ measured at different values of the bandwidth parameter. Vertical axis: difference between bandwidths of 3 and 4, normalized to the $\mathcal{N}_c^\mathrm{bw4}$. Horizontal axis: same for bandwidths 5 and 4. The dashed lines mark the 16- and 84- percentiles containing 68\% of the distribution of $\Delta_{rel}^{3, 4}$ and $\Delta_{rel}^{5, 4}$.}
     \label{fig:rel_Nc_errs}
 \end{figure}

Figure \ref{fig:rel_Nc_errs} shows the joint distribution of $\Delta_{rel}^{4, 3}$ and $\Delta_{rel}^{5, 4}$ for the map at \texttt{nside=32}. The majority of pixels show relative differences of $< 0.1$ (10\%). 
The 16--percentile and 84--percentiles of the distribution of $\Delta_{rel}^{4, 3}$ are -0.3 and 0.01, respectively. For $\Delta_{rel}^{5, 4}$, these values are -0.1 and 0.01. There are long tails extending out to -2 for the distribution of $\Delta_{rel}^{4, 3}$ and -0.6 for that of $\Delta_{rel}^{5, 4}$. We interpret these distributions as follows: by using a smaller bandwidth, a larger number of clouds is detected (due to the increased velocity resolution). The bulk of the observed differences (within 10---30\% relative difference) are likely related to changes in the identification of low-column density clouds. These clouds are prevalent in the HI decomposition, as discussed in \citet{Panopoulou2020}, and contribute a low-level noise to the determination of $\mathcal{N}_c$. We will assign this low-level noise as a floor in the uncertainty on $\mathcal{N}_c$. The longer (negative) tails of the distributions are likely related to edge cases, where clouds of significant column density were unresolved at low bandwidth but become resolved with a small increase in the velocity resolution.

Because our clustering algorithm deals with Gaussian uncertainties, we need to translate these asymmetric differences into an equivalent standard deviation. We choose to err on the side of overestimating the uncertainties, by adopting the following estimate of the uncertainty on $\mathcal{N}_c$: 
\begin{equation}
\sigma(\mathcal{N}_c) = \max \left(0.3,\,  \Delta_{rel}^{3, 4},\, \Delta_{rel}^{5, 4}  \right),
\label{eq:nc_err}
\end{equation}
meaning that we assign the maximum observed difference between different bandwidth runs as the (symmetric) $\rm 1-\sigma$ uncertainty in each pixel. The aforementioned low-level noise contributed by faint clouds represents the uncertainty floor that we assign (and corresponds to the absolute value of the 16-percentile of $\Delta_{rel}^{3, 4})$.

\bsp	
\label{lastpage}
\end{document}